\documentclass[preprint]{aastex}
\usepackage{hyperref}

\shorttitle{Differential Rotation on II Pegasi}
\shortauthors{Roettenbacher et al.}

\begin{document}

\title{A Study of Differential Rotation on II Pegasi via Photometric Starspot
Imaging}

\author{Rachael M.\ Roettenbacher\altaffilmark{1}, Robert O.\ Harmon, Nalin
Vutisalchavakul\altaffilmark{2}}
\affil{Department of Physics and Astronomy,}
\affil{Ohio Wesleyan University, Delaware, OH 43015}
\email{roharmon@owu.edu}
\altaffiltext{1}{Presently at Department of Physics, Lehigh University,
16 Memorial Drive E., Bethlehem, PA 18015}
\altaffiltext{2}{Presently at Department of Astronomy, The University of Texas
at Austin, Austin, TX 78713}

\and

\author{Gregory W. Henry}
\affil{Center of Excellence in Informations Systems,}
\affil{Tennessee State University,}
\affil{3500 John A. Merritt Blvd., Box 9501, Nashville, TN 37209}

\begin{abstract}
We present the results of a study of differential
rotation on the K2 IV primary of the RS CVn binary II Pegasi (HD 224085) 
performed by inverting light curves to produce images of the dark starspots 
on its surface. The data
were obtained in the standard Johnson $B$ and $V$ filter passbands via the
Tennessee State University T3 0.4-m Automated Photometric Telescope
from JD 2447115.8086 to 2454136.6221 (1987 November 16 to 2007 February 5). 
The observations were subdivided into 68 data sets consisting of pairs of 
$B$ and $V$ light curves, which
were then inverted using a constrained non-linear inversion algorithm that
makes no \emph{a priori} assumptions regarding the number of spots or their
shapes. The resulting surface images were then assigned to 21 groups
corresponding to time intervals over which we could observe the evolution of
a given group of spots (except for three groups consisting of single data sets).
Of these 21 groups, six showed convincing evidence of differential rotation over
time intervals of several months. For the others, the spot configuration was such that
differential rotation was neither exhibited nor contraindicated. The differential rotation 
we infer is in the same sense as that on the Sun: lower latitudes have shorter 
rotation periods. From plots of the range in longitude spanned by the spotted regions vs.\ 
time, we obtain estimates of the differential rotation coefficient $k$ defined in
earlier work by Henry et al., and show that our results for its value are consistent with
the value obtained therein.
\end{abstract}

\keywords{starspots---stars: activity---stars: imaging---stars: individual (II Pegasi)---
stars: variables: general---binaries: close} 

\section{\label{sec:Introduction} Introduction}

II Pegasi is an SB1 binary system for which the primary component
was determined to be of spectral class K2-3 IV-V by \citet{Rucinski1977}. It was classified
as an RS CVn system by \citet{Vogt1981a}. The first photometric light curves were obtained
by \citet{Chugainov1976}, who found variability with a period of approximately 6.75~d and
interpreted the asymmetric light curve in terms of rotational modulation due to large, 
cool starspots. In 1986 September, the difference between maximum and minimum light for the 
$V$ filter reached 0.5~mag, implying a projected spot area coverage of the visible hemisphere at minimum light on the order of 50\% \citep{Doyle1989}. 

On the basis of high-quality radial velocity measurements,
\citet{Berdyugina1998a} 
determined the revolution period of the binary to be $6.724333 \pm 0.000010$~d.
The same authors performed a detailed model atmosphere analysis of high-resolution and
high signal-to-noise CCD spectra, obtaining values for the photospheric temperature and 
surface gravity of the primary star of $T_\mathrm{eff} = 4600$~K and $\log g = 3.2$, 
with $g$ expressed
in cgs  units. These values correspond to a K2 IV star of mass 
$M = 0.8 \pm 0.1\mbox{ M}_\odot$. 
They estimated the radius of the primary as $R = 3.4 \pm 0.2\mbox{ R}_\odot$
and the
inclination to be $\alpha = 60^\circ \pm 10^\circ$ on the assumption that the
rotational axis 
is perpendicular to the orbital plane. Based on the fact that the secondary star
is unseen at all wavelengths and thus has luminosity at least 100 times smaller
than that of the primary, they estimated the secondary to
be an M0-M3 red dwarf. 

II Peg is among the most active RS CVn systems, and it is one of a small number
of binaries in which the H$\alpha$ line is always seen in emission
\citep{NationsRamsey1981}.
Recently, \citet{Frasca2008} reported on contemporaneous photometric and
spectroscopic observations of II Peg, finding that the H$\alpha$ emission and
photometric intensity are strongly anticorrelated, suggesting that regions of
high chromospheric activity are physically associated with the spots. This
conclusion was corroborated by a rotational modulation of the intensity of the
He \textsc{I} D$_3$ line. Based on an estimated radius of 
$R = 2.76\mbox{ R}_\odot$ and $v\sin i = 22.6\mbox{ km s$^{-1}$}$, they
estimated 
the inclination between the rotation axis and the line of sight to be
$\alpha = 60^\circ$$^{+30}_{-10}$. \citet{Messina2008} confirmed via long-term
monitoring of $V$ as well as the $B-V$ and $U-B$ colors that II Peg is redder when it
is dimmer, as would be expected if the dimming is caused by cool spots.

A number of studies have attempted to determine the spot temperatures. \citet{Vogt1981b}
modeled light and color curves obtained in 1977 with a single circular spot, finding a spot
temperature of $T_\mathrm{spot} = 3400 \pm 100\mbox{ K}$. 
\citet{NationsRamsey1981} obtained
$T_\mathrm{spot} = 3600\mbox{ K}$ from observations in the Fall of 1979;
\citet{PoeEaton1985} obtained 
$T_\mathrm{spot} = 3620\mbox{ K}$ for Fall 1980; \citet{Rodono1986} obtained 
$T_\mathrm{spot} = 3300\mbox{ K}$ for Fall 1981; \citet{ByrneMarang1987} obtained
$T_\mathrm{spot} = 3700\mbox{ K}$ for Fall 1986; and \citet{Boyd1987} obtained 
$T_\mathrm{spot} = 3450\mbox{ K}$ for 1986--1987. By modeling the strengths of TiO
absorption bands, \citet{ONeal1998} found evidence for multiple spot temperatures, finding
$T_\mathrm{spot}$ to vary between $3350 \pm 60\mbox{ K}$ and $3550 \pm 70\mbox{
K}$ as the star
was observed through slightly less than one rotational period. More recently, from a spot 
model applied to contemporaneous photometry and spectroscopy, \citet{Frasca2008} obtained 
$T_\mathrm{spot} \approx 3600\mbox{ K}$.

\citet{Henryetal1995} used a simple analytic two-spot model to fit photoelectric light curves 
of four chromospherically active binaries: $\lambda$ And, $\sigma$ Gem, V711 Tau,
and II Peg. The II Peg data were acquired
from 1973--1992, and subdivided into 37 individual light curves. 
They plotted ``migration curves'' for twelve long-lived spots they identified in
the data from the times of minimum light obtained via the spot-model curve fits. A migration
curve shows the variation in the phase of minimum light with time, where the phase was
computed using the orbital ephemeris and represents the fractional part of the number of 
rotation periods since an arbitrary starting time. Assuming tidal locking and
no differential rotation, a given spot would always
cross the central meridian of the stellar disc as seen from Earth once per
revolution period and thus always at the same phase. However, if the
star exhibits latitude-dependent differential rotation, we would expect to see a
given spot progressively advanced or retarded in phase relative to the
orbital ephemeris. A plot of the phase of minimum light versus time for a given
spot should then be a straight line with slope determined by the difference in
the rotation period of the latitude of the spot and the revolution period. This
was precisely what Henry 
et al.\ observed in the data, for II Peg and the other stars. The plots for
different spots
had different slopes, demonstrating latitude-dependent differential rotation. The degree of 
differential rotation was specified in terms of the differential rotation
coefficient, $k$,
defined for the Sun by fitting the rotation period as a function of latitude with the relation
\begin{equation}
P(\theta) = \frac{P_\mathrm{eq}}{1 - k\sin^2\theta},
\label{eq:DiffRotP}
\end{equation}
where $P(\theta)$ is the rotation period for latitude $\theta$ and
$P_\mathrm{eq}$ is the
rotation period at the equator. For the Sun, $k = 0.19$. If the differential rotation of other 
stars has the same functional
form as for the Sun, and if the rotation periods for spots sampling a range of latitudes are
determined for a star, then the coefficient $k$ is given by
\begin{equation}
\frac{P_\mathrm{max} - P_\mathrm{min}}{P_\mathrm{avg}} = kf,
\label{eq:kf}
\end{equation}
where $P_\mathrm{max}$, $P_\mathrm{min}$, and $P_\mathrm{avg}$ are the maximum,
minimum, and
average observed periods, and $f$ is a distribution function which relates the 
the total range in rotational period sampled to the number of spots for which the
period has been determined \citep{HallHenry1994}. The value of $f$ ranges from 0.5 for 
two spots to over 0.9 when the number of spots exceeds
six. \citet{Henryetal1995} used
eight of the twelve spots they observed on II Peg (four spots were observed over
intervals too short
to allow their periods to be obtained reliably) to determine $k$ using equation
(\ref{eq:kf}), with the result $k = 0.005 \pm 0.001$. 

\citet{Rodono2000} performed an analysis similar to the present study,
inverting light curves acquired between 1974 and 1998 to produce images
of the stellar surface. In contrast to the smoothing function used here (see
\S \ref{sec:LI_Algorithm} and in particular equation
(\ref{eq:SmoothingFunction})), they used maximum 
entropy and Tikhonov regularization. They concluded that the 
distribution of spots on II Peg consists of a component distributed
uniformly in longitude which does not rotationally modulate the light curve
(but does produce a secular variation in the mean intensity), plus an
unevenly distributed component responsible for the rotational modulation. Their analysis
indicated that the uniformly distributed component varied in total area
with a period of $\sim 13.5\mbox{ yr}$. They determined the unevenly
distributed component to be concentrated around three active longitudes, one of
these having an
essentially permanent presence but a cycle in spot area with period $\sim
9.5\mbox{ yr}$.
They found the activity of the other two
active longitudes to switch back and forth, with one active while the other is
inactive, with a period of $\sim 6.8\mbox{ yr}$.
However, there is an interval of $\sim 1.05\mbox{ yr}$ before the switch in which
both longitudes are
active. There is thus a period of $\sim 6.8 - 2(1.05) = 4.7\mbox{ yr}$ during which
only one of the two longitudes engaged in the ``flip-flop'' behavior is active, which agrees
with the switching period deduced by \citet{BerdyuginaTuominen1998} from the times of 
light minima. 

From a periodogram analysis, \citet{Henryetal1995} found periodicities in the mean magnitudes 
for the spot-model fits of their 37 light curves of $4.4 \pm 0.2\mbox{ yr}$ and 
$11 \pm 2\mbox{ yr}$. They interpreted the 4.4-yr period as reflecting the
average lifetime of the spots and the 11-yr period as representing a different timescale.
Rodon\`{o} et al.\ interpreted the 4.7-yr periodicity arising in their analysis as 
corresponding to the 4.4-yr period obtained by Henry et al., while they interpreted
the 9.5-yr period they saw in the total area of the spot component which is unevenly 
distributed in longitude as corresponding to the 11-yr period found by Henry et al. 

Henry et al.\ noted that their two-spot model, which assumed circular spots
varying only in radius over time, was not fully adequate to explain the
variations with time of 
the II Peg light curve. In particular, when the amplitude of the rotational
modulation due 
to spots they designated G and H was diminishing, the mean brightness of the star stayed 
roughly constant. Similar behavior was seen for another pair of spots, which they designated
J and K. If the decrease in amplitude were due simply to a decrease in the spot radii, the mean
brightness of the star should have increased (assuming no change in the brightness of the
photosphere outside the spots). On the other hand, if instead the spots were being drawn out
in longitude by differential rotation while maintaining nearly constant area, then the
amplitude would decrease while the mean brightness stayed constant, as observed. 

The present study is suited to look for evidence for such drawing out 
(or compression) in longitude of active regions by differential rotation,
as we produce images of the active regions and observe changes in them over periods of
several months. By simultaneously inverting contemporaneous $B$ and $V$ light curves,
we exploit differences in the limb darkening as seen through the two filters to achieve
significantly better latitude resolution than is possible when using light curves obtained
through only a single filter \citep{HarmonCrews2000}, thereby allowing us to directly
detect differential rotation in our images. It should be noted, however, that we do not
claim to obtain accurate spot latitudes from our two-filter inversions; 
nonetheless, simulations
like those detailed in Harmon \& Crews show that \emph{relative} spot latitudes can be
obtained with good reliability, i.e., when two spots are present, the one at the lower
latitude is rendered as such. In this regard our approach differs from that of
Rodon\`{o} et al., who used their $V$-filter surface imagery just to derive quantities that they claim are independent of the regularization criterion, 
such as the distributions of
the spots versus longitude, the changes in the distribution over time, and the
variations of the total area covered by spots. 

In \S \ref{sec:LI_Algorithm}, we discuss the method used to invert
the light curves so as to produce these images. In \S \ref{sec:DataAnalysis}, we
discuss the division of the over nineteen years worth of $B$ and $V$ 
light curves into separate data sets
and the procedure used to process them for inversion. In \S \ref{sec:Results},
we discuss
in detail the results for the six intervals over which we saw good evidence in our images for
alteration of the active region configuration by latitude-dependent differential rotation.
Finally, in \S \ref{sec:Discussion}, we show that our results are consistent with the
result for the value of the differential rotation coefficient $k$ for II Peg
inferred by \citet{Henryetal1995} based
on longer-term monitoring of the times of minimum light due to individual spots using
their two-spot model rather than surface imaging. 

\section{\label{sec:LI_Algorithm} The Light-curve Inversion Algorithm}

Light-curve Inversion (LI) is a photometric imaging technique which produces a
map of a star's surface based on the brightness variations produced as dark
(or possibly bright) starspots are carried into and out of view of Earth by the
star's rotation. It makes no \emph{a priori} assumptions regarding the number
of spots on the surface or their shapes. The details regarding the
implementation of the algorithm are presented in \citet{HarmonCrews2000}, 
along with the results of extensive tests in which
artificial stellar surfaces were used to create light curves, which were then
inverted. In Harmon \& Crews, the technique is called ``Matrix Light-curve Inversion,'' 
because it evolved from the original formulation described in \citet{Wild1989} and called
by that name. However, because the formulation described in
\citet{HarmonCrews2000} and as modified in the present work no longer uses matrices, we shortened
the name. Here we outline the method, and refer the reader to Harmon \& Crews for more details.

The stellar surface is subdivided into $N$ bands in latitude of equal
angular widths $\Delta \theta = \pi/N$. Each latitude band is further subdivided
into patches which are all ``spherical rectangles'' of equal widths in
longitude $\Delta \phi = 2\pi/M_i$, where $M_i$ is the number of patches in the
$i^\mathrm{th}$ latitude band. The $M_i$ are chosen to be proportional to the
cosine of the latitude (to within the constraint that the $M_i$ must be
integers) so that the areas of all the patches are nearly equal. The visible
pole is defined to be the north pole, with latitude $+90^\circ$, while the
hidden south pole has latitude $-90^\circ$. The $j^\mathrm{th}$ patch in the
$i^\mathrm{th}$ latitude band is designated patch $(i,j)$. The first patch in
each latitude band, patch $(i,1)$, straddles the meridian with longitude $\phi =
0$, defined to be the one which
intersects the equator on the approaching limb of the star at an arbitrarily
chosen reference time $t_0$. Longitude increases in the direction of the star's
rotation, so the sub-observer longitude at $t = t_0$ is thus
$90^\circ$.
In the absence of interstellar absorption, at the time $t_{nk}$ of observation
number $k$ through filter $n$, the intensity $I_{nk}$ observed at Earth is 
(in the limit that the number of patches is large)
\begin{equation}
I_{nk}\index{} = \sum_{i=1}^{N_s}\sum_{j=1}^{M_i}
\Omega_{nk;ij}L_{nk;ij}J_{n;ij}, \qquad n = 1, \ldots, Q, \quad k = 1, \ldots,
P_n,
\label{eq:Ink}
\end{equation}
where $Q$ is the number of filters, $P_n$ is the number of observations
through filter $n$, $J_{n;ij}$ is the specific intensity (W m$^{-2}$
sr$^{-1}$) along the
outward normal of patch $(i,j)$ integrated over the passband of filter $n$,
$\Omega_{nk;ij}$ is the solid angle of patch
$(i,j)$ as seen from Earth at time $t_{nk}$ (we set $\Omega_{k;ij} = 0$ if the
patch is on the far side of the star), $L_{nk;ij}$ is the factor by which the
specific intensity emitted in the direction of Earth is attenuated by limb
darkening compared to that emitted along the outward normal (so that
$L_{nk;ij}J_{n;ij}$ is the specific intensity emitted along the line of sight to
Earth), and $N_s$ is the index of the southernmost latitude band which is
visible from Earth. 

The goal of LI is to find a set of computed patch intensities $\hat{J}_{n;ij}$
that mimics the actual variations of surface brightness across the stellar
surface as closely as possible. (We use a caret over a
quantity to indicate that it represents a value as computed by the LI
algorithm.) Since we generally do not know the actual radius
and distance of the star very precisely, we content ourselves with finding only
the \emph{relative} brightnesses of the patches to one another. To this end we
simply define the radius of the star to be 1 and use the area of a
patch projected onto the plane of the sky as a proxy for the solid angle it subtends at
Earth. 

We use the limb-darkening coefficients
published by \citet{VanHamme1993} to determine the values of the $L_{nk;ij}$ in
equation (\ref{eq:Ink}). The benefit of observing through multiple filters is
that we can take advantage of the differences in the degree of limb darkening as
seen through different filters in order to significantly increase the latitude
resolution of the
inversions, as explained in \citet{HarmonCrews2000}. In order to take advantage
of this
information, we must simultaneously invert all of the filter light curves. This
in
turn requires that we couple together the
$\hat{J}_{n;ij}$ for different values of the filter index, $n$. To do this, we
designate the
filter for which the light curve has the lowest noise as
the ``primary filter'' and assign it filter index $n = 1$. For simplicity we
assume that the actual
stellar surface can be described via a two-component model in which all the
spots have the same temperature $T_\mathrm{spot}$ and thus emit the same
specific intensity along the outward normal as seen through filter $n$, which we
designate at $J_{n;\mathrm{spot}}$; similarly, we assume that all points on the
surface outside spots are part of a photosphere of uniform temperature
$T_\mathrm{phot}$ and emitting specific intensity $J_{n;\mathrm{phot}}$ along
the outward normal. However, it should be noted that the reconstructed surface
created by the inversion of the data does \emph{not} have the property that the
$\hat{J}_{n;ij}$ can have only one of two values; they are continuous
variables.
We then define the intensities of the patches as viewed through filter $n \neq
1$ via the linear scaling 
\begin{equation}
\hat{J}_{n;ij} \equiv \frac{r_n}{1 - s_1}\left[(s_n -
s_1)\hat{J}_{1;\mathrm{avg}} + (1 - s_n)\hat{J}_{1;ij}\right].
\label{eq:Scaling}
\end{equation}
Here $\hat{J}_{1;\mathrm{avg}}$ is the average value of the $\hat{J}_{1;{ij}}$,
$r_n$ is the estimated value of $J_{n;\mathrm{phot}}/J_{1;\mathrm{phot}}$ on
the actual stellar surface, and $s_n$ is the estimated value of
$J_{n;\mathrm{spot}}/J_{n;\mathrm{phot}}$ for the actual stellar surface. We 
estimate these ratios by calculating the Planck function at the central
wavelength of the filter in question at the assumed spot and
photosphere temperatures. 
It would be more accurate to integrate the product of the Planck function
and the filter sensitivity functions over
their passbands, but typically we do not know the spot and photosphere
temperatures with sufficient precision to justify the extra effort.
The scaling given in equation (\ref{eq:Scaling}) from $\hat{J}_{1;ij}$ to 
$\hat{J}_{n;ij}$ has the property that when 
$\hat{J}_{1;ij}/\hat{J}_{1;\mathrm{avg}} =
J_{1;\mathrm{spot}}/J_{1;\mathrm{phot}}$ according to our estimate, then
$\hat{J}_{n;ij}/\hat{J}_{n;\mathrm{avg}} =
J_{n;\mathrm{spot}}/J_{n;\mathrm{phot}}$ as well. We are using
$\hat{J}_{1;\mathrm{avg}}$ as a proxy for $J_{1;\mathrm{phot}}$, which should be
a reasonable approximation as long as the stellar photosphere comprises most of
the surface.

Since the patch brightnesses through all the filters in the model are entirely
determined in terms of their brightnesses $\hat{J}_{1;ij}$ as seen through filter 1, 
the problem reduces to finding these values, so for notational simplicity 
we define $\hat{J}_{ij} \equiv \hat{J}_{1;ij}$. As is well known,
the problem of determining the $\hat{J}_{ij}$ is extremely sensitive to
the presence of even small amounts of noise in the data. This can be seen by
considering the effect on the light curve produced by a myriad of small spots
distributed all over the surface. As the star rotated, at any given time nearly
equal numbers of spots would be rising over the approaching limb and setting
over the receding limb. Thus, the total contribution to the star's brightness
from the spots would be nearly but not exactly constant, so that the effect of
the spots would be to impart a small-amplitude, high-frequency ripple on the
light curve, very similar to the effect of random noise in the observations.
Conversely, if we attempt to fit noisy data, then unless precautions are taken,
the resulting model surface will be
covered with spurious small spots introduced in order to ``explain'' the
presence of the noise in the signal.

To avoid this dilemma, rather than simply finding the set of $\hat{J}_{ij}$
that yields the best fit to the light curve data, we determine the
$\hat{J}_{ij}$ by finding the set of them which 
minimizes the \emph{objective function} \citep{Twomey1977,CraigBrown1986}:
\begin{equation}
E(\mathbf{\hat{J}},\mathbf{I},\lambda,B) = G(\mathbf{\hat{J}},\mathbf{I})
+ \lambda S(\mathbf{\hat{J}},B).
\label{eq:ObjectiveFunction}
\end{equation}
Here $\mathbf{\hat{J}}$ represents the set of the $\hat{J}_{ij}$, while
$\mathbf{I}$ represents the set of
observed intensities $I_{nk}$, i.e., the data light curve.
The function $G(\mathbf{\hat{J}},\mathbf{I})$ expresses the goodness-of-fit of
the calculated light curve $\mathbf{\hat{I}}$ (with components $\hat{I}_{nk}$)
obtained from $\mathbf{\hat{J}}$ to the data light curve $\mathbf{I}$, such that
smaller values of $G(\mathbf{\hat{J}},\mathbf{I})$ imply a better fit to the
data. The \emph{smoothing function} $S(\mathbf{\hat{J}},B)$
is defined such that it takes on smaller values for surfaces
that are ``smoother'' in some appropriately defined sense. Finally, $\lambda$
is an adjustable Lagrange multiplier called the \emph{tradeoff parameter}, and
$B$ is an adjustable parameter called the \emph{bias parameter}, which is
discussed below. Note that as $\lambda
\rightarrow 0$, the first term on the right dominates, so that minimizing $E$ is
equivalent to minimizing $G$, and we obtain the solution which best fits the
light curve data but suffers from the spurious noise artifacts discussed above.
On the other hand, as $\lambda \rightarrow \infty$, the second term dominates,
so that
minimizing $E$ produces a very ``smooth'' surface lacking in noise artifacts,
but also producing a very poor fit to the data. Thus, by varying $\lambda$, we
adjust the tradeoff between goodness-of-fit and smoothness of the model
surface. If we choose $\lambda$ such that $G(\mathbf{\hat{J}},\mathbf{I})$ is
equal to a corresponding estimate of the amount of noise in the data, then in a
rough
sense we can say that by minimizing the objective function, we find the
smoothest solution $\mathbf{\hat{J}}$ for which the corresponding light curve
$\mathbf{\hat{I}}$ fits the data light curve $\mathbf{I}$ to a degree which is 
as good as but not better than
is justified by the noise in the data. In this way, we obtain a model
surface which fits the data well, but not so well that it is dominated by noise
artifacts.

For the goodness-of-fit function in this study, we use
\begin{equation}
G(\mathbf{\hat{J}},\mathbf{I}) = \frac{(2.5 \log_{10}e)^2}{P}\sum_{n=1}^Q
\frac{1}{\sigma_n^2}
\sum_{k=1}^{P_n}\left(\frac{I_{nk}-\hat{I}_{nk}}{I_{nk}}\right)^2.
\label{eq:GoodnessOfFitFunction}
\end{equation}
Here we assume that light curves have been obtained through $Q$
different photometric filters ($Q = 2$ in the present work since we use $B$- 
and $V$-filter data), and that the magnitudes have been converted to
intensities. Since our goal is only to find the relative
values of the $\hat{J}_{ij}$, it suffices to use relative rather than absolute
intensities for the light curve data in the calculation of
$G(\mathbf{\hat{J}},\mathbf{I})$. The number of observations in the
light curve obtained through filter $n$ is $P_n$, while $P = \sum_n P_n$ is
the total number of data points in all the light curves. The estimated noise
variance in the light curve data for filter $n$ expressed in magnitudes is
$\sigma_n^2$. In \citet{HarmonCrews2000} it is shown that to a good
approximation, the true noise variance $\tilde{\sigma}_n^2$ is given by
\begin{equation}
\tilde{\sigma}_n^2 \approx \frac{(2.5 \log_{10}e)^2}{P_n}
\sum_{k=1}^{P_n}\left(\frac{I_{nk}-\tilde{I}_{nk}}{\tilde{I}_{nk}}
\right)^2 ,
\end{equation}
where $\tilde{I}_{nk}$ is the true noise-free value of the intensity
(which is of course unknown unless one is doing a simulation). If
we define $\epsilon_{nk} \equiv I_{nk} - \tilde{I}_{nk}$ to be the true error
in the measurement $I_{nk}$, and $\delta_{nk}
\equiv \hat{I}_{nk} - \tilde{I}_{nk}$ to be the deviation between the calculated
 and true intensities, then with this notation
\begin{equation}
\frac{I_{nk} - \tilde{I}_{nk}}{\tilde{I}_{nk}} =
\frac{\epsilon_{nk}}{\tilde{I}_{nk}},
\end{equation}
while
\begin{equation}
\frac{I_{nk} - \hat{I}_{nk}}{I_{nk}} = \frac{\epsilon_{nk} -
\delta_{nk}}{\tilde{I}_{nk} + \epsilon_{nk}}.
\label{eq:Gterm}
\end{equation}
If $\mathbf{\hat{I}}$ is a good match to the data $\mathbf{I}$, then the
$\epsilon_{nk}$ and $\delta_{nk}$ are small quantities, and we can expand
the right side of equation (\ref{eq:Gterm}) as
\begin{equation}
\frac{I_{nk} - \hat{I}_{nk}}{I_{nk}} =
\frac{\epsilon_{nk}-\delta_{nk}}{\tilde{I}_{nk}}\left(1 -
\frac{\epsilon_{nk}}{\tilde{I}_{nk}} + \ldots\right)
= \frac{\epsilon_{nk} - \delta_{nk}}{\tilde{I}_{nk}} +
\ldots
\end{equation}
Then
\begin{eqnarray}
G(\mathbf{\hat{J}},\mathbf{I}) &=& \frac{1}{P} \sum_{n=1}^Q
\frac{(2.5 \log_{10}e)^2}{\sigma_n^2} \sum_{k=1}^{P_n} 
\left[\frac{\epsilon_{nk}^2}{\tilde{I}_{nk}^2} +
\frac{\delta_{nk}^2 - 2\epsilon_{nk}\delta_{nk}}{\tilde{I}_{nk}^2} + \ldots
\right] \nonumber \\
&=& \frac{1}{P}\sum_{n=1}^{P_n}\left[P_n\frac{\tilde{\sigma}_n^2}{\sigma_n^2}
+ \frac{(2.5\log_{10}e)^2}{\sigma_n^2} \sum_{k=1}^{P_n} \frac{\delta_{nk}^2 -
2\epsilon_{nk}\delta_{nk}}{\tilde{I}_{nk}^2} + \ldots \right]
\end{eqnarray}

If the reconstructed intensities $\hat{I}_{nk}$ perfectly matched the
true intensities $\tilde{I}_{nk}$, and in addition the estimated noise
variances $\sigma_n$ were equal to the true noise variance $\tilde{\sigma_n}$,
then we would have all $\delta_{nk} = 0$, and so we would have
$G(\mathbf{\hat{J}},\mathbf{I}) = 1$ to lowest order in the $\epsilon_{nk}$.
For this reason, given the estimates $\sigma_n$ and for a given value of the bias
parameter $B$, we vary $\lambda$ until the stopping
criterion $G(\mathbf{\hat{J}},\mathbf{I}) = 1$ is attained to a predetermined
precision.

The smoothing function used in this study is
\begin{equation}
S(\mathbf{\hat{J}},B) = \sum_{i=1}^N \sum_{j=1}^{M_i} c_{ij}(\hat{J}_{ij} -
\hat{J}_\mathrm{avg})^2,
\label{eq:SmoothingFunction}
\end{equation}
where $c_{ij} = 1$ if $\hat{J}_{ij} \le \hat{J}_\mathrm{avg}$, while $c_{ij} =
B$ if $\hat{J}_{ij} > \hat{J}_\mathrm{avg}$. Thus, patches brighter
or darker than average incur a penalty in that they increase the value of
$S(\mathbf{\hat{J}},B)$ (and thus the function 
$E(\mathbf{\hat{J}},\mathbf{I},\lambda,B)$ to be minimized) 
by an increasing amount as the deviation from the
average increases. Note that $S(\mathbf{\hat{J}},B)$ satisfies the criterion
that it takes on its minimum possible value of zero for a featureless surface
which is perfectly ``smooth'' in that all the patch brightnesses $\hat{J}_{ij}$
are equal, and that surfaces showing greater deviations about the average are
judged as ``rougher.'' For $B > 0$, the penalty for a patch being brighter than
average by a given amount is $B$ times larger than for a patch darker
than average to the same degree. Thus, $B$ biases the solution toward having
most patches just slightly brighter than average to represent the stellar
photosphere, which is assumed to be almost uniformly bright like the Sun's,
while a smaller number are much darker than average to represent the dark
starspots. This is the reason for the name ``bias parameter.'' 

The simulations described in \citet{HarmonCrews2000} show that as $B$ is increased, the ratio
$\min(\hat{J}_{ij})/\hat{J}_\mathrm{avg}$ decreases, so that the darkest patch
becomes darker relative to the average patch brightness. We use this ratio as a
proxy for the assumed ratio of the spot and photosphere brightnesses as seen
through filter 1. For
a given value of the tradeoff parameter $\lambda$, the bias parameter $B$ can be
varied until $\min(\hat{J}_{ij})/\hat{J}_\mathrm{avg} = s_1$, the estimated
spot-to-photosphere brightness ratio seen through filter 1, to a
predetermined
precision. The scaling given by equation (\ref{eq:Scaling}) ensures that
$\min(\hat{J}_{n;ij})/\hat{J}_{n;\mathrm{avg}} = s_n$, the estimated
spot-to-photosphere brightness ratio for filter $n$.

The procedure for inverting a series of light curves obtained through
a set of filter passbands is then as follows. The input parameters are the
estimated noise variances $\sigma_n^2$ of the light curves, the estimated
spot and photosphere temperatures $T_\mathrm{spot}$ and $T_\mathrm{phot}$,
and the inclination angle $\alpha$ of the rotation axis to the line of sight. 
The $\sigma_n$ in the definition of
$G(\mathbf{\hat{J}},\mathbf{I})$, $T_\mathrm{spot}$, and $T_\mathrm{phot}$ are 
used to obtain the values
of $s_n$ and $r_n$ (including $n=1$) in equation (\ref{eq:Scaling}), and
the inclination $\alpha$ is used in finding $\Omega_{nk;ij}$ and $L_{nk;ij}$ in 
equation (\ref{eq:Ink}). 
As described in \citet{HarmonCrews2000}, two copies of a root-finding subroutine
are used in concert so as to find the values of $\lambda$ and $B$ such that
$G(\mathbf{\hat{J}},\mathbf{I}) = 1$ \emph{and}
$\min(\hat{J}_{n;ij})/\hat{J}_{n;\mathrm{avg}} = s_n$ to the desired precision.

The result is a set of solutions, one for each combination of the input
parameters. How we select one of these to represent the ``best'' solution is
described in \S \ref{sec:DataAnalysis}.

\section{\label{sec:DataAnalysis} Data Analysis}

\begin{figure}[!htbp]
\centering
\epsscale{0.8}
\plotone{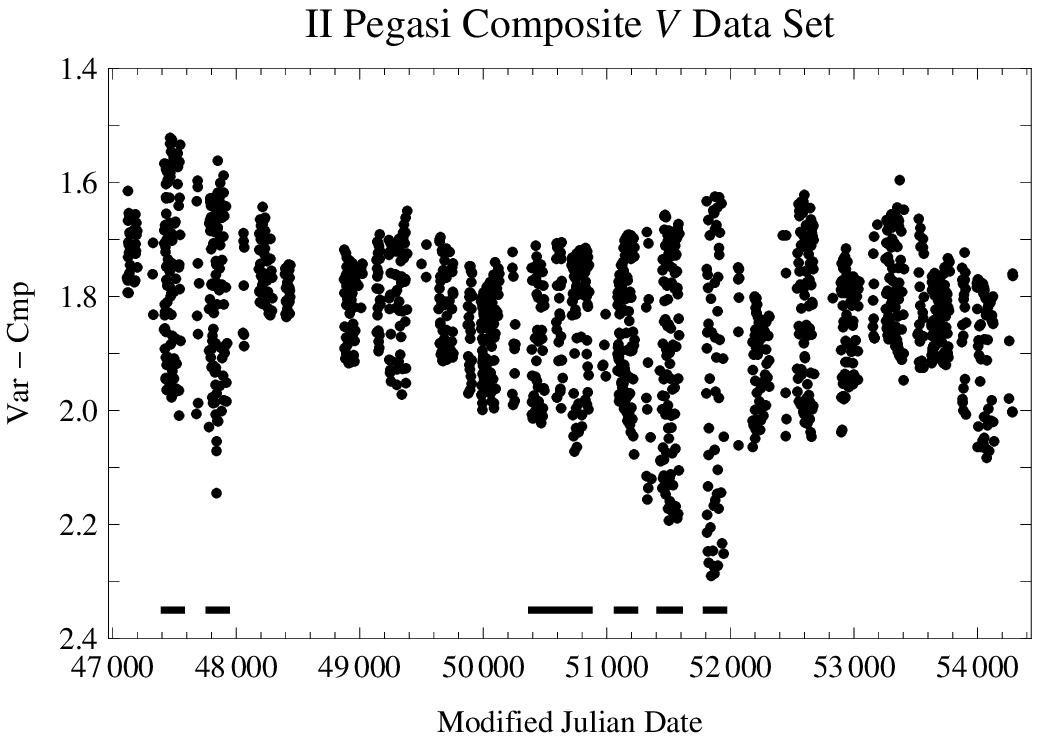}
\vspace{12pt}
\plotone{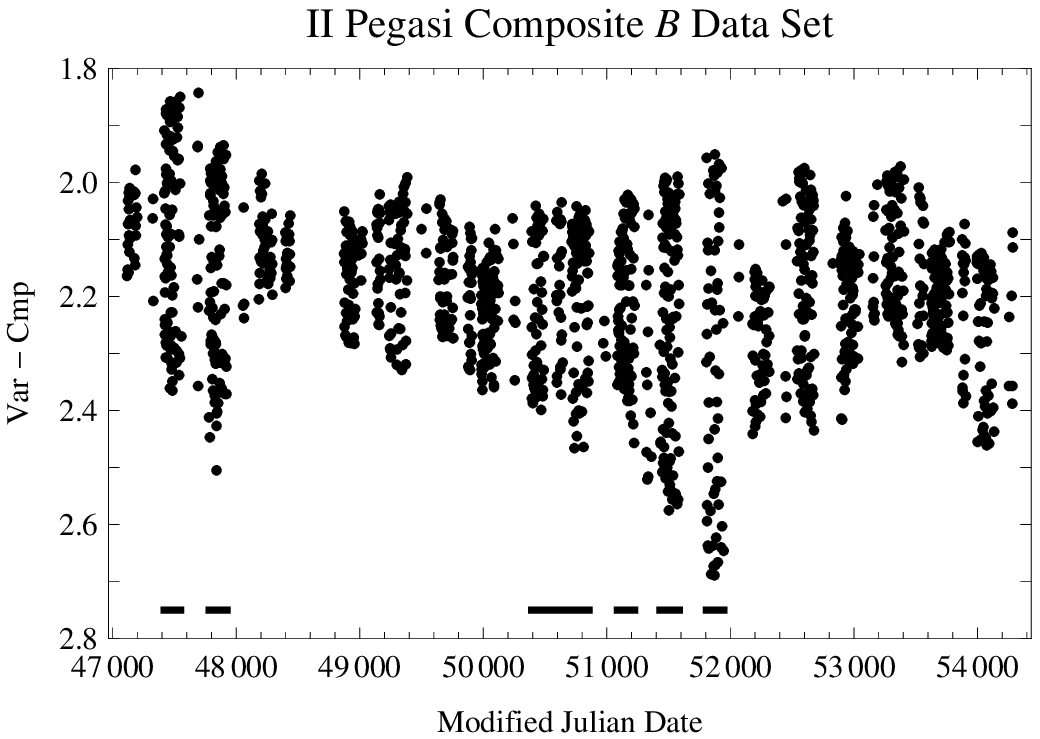}
\epsscale{1}
\caption{\label{fig:CompleteDataSet} The complete $B$ (top) and $V$ (bottom) data sets,
plotted as the difference between the magnitudes of II Pegasi and the comparison star
(Var$-$Cmp).
The horizontal bars indicate the six time intervals for which our analysis produced good
evidence for differential rotation.}
\end{figure}

The raw data consisted of Johnson $B$ and $V$ differential magnitudes paired
with corresponding heliocentric Modified Julian Dates, acquired from
heliocentric MJD 47115.8086 to 54136.6221 (1987 November 16 to 2007 February 5)
with the 
Tennessee State University T3 0.4-m Automated Photometric Telescope 
at Fairborn Observatory in Arizona \citep{Henry1995a,Henry1995b}. The complete $B$ 
and $V$ data sets are plotted as the upper and lower panels of Figure 
\ref{fig:CompleteDataSet}.

The first task was to convert the Modified Julian Dates to rotational phases. The
rotational phase $\Phi$ is defined as
\begin{equation}
\Phi(t) = \frac{t - t_0}{T} - \left\lfloor\frac{t - t_0}{T}\right\rfloor,
\end{equation}
where $t$ is the time of the observation, $t_0$ is an arbitrary reference time
used for all the observations, $T$ is the rotational period of the star, and
$\lfloor x\rfloor$ is the greatest integer which is less than or equal to $x$.
Thus, $\Phi(t)$ represents the fraction of a rotation through which the star
has turned relative to the orientation it had at time $t_0$. 
On the assumption that the star exhibits
differential rotation, there is no such thing as \emph{the} rotational period,
so $T$ here represents a suitable average of the rotation period over all
latitudes. On the assumption of tidal locking, it is reasonable to use
the orbital period for this average. In the present study, 
$T = 6.724333$~d was used based on the
orbital period obtained by \citet{Berdyugina1998a}, and $t_0$ was
chosen as JD 2443033.47, based on the orbital ephemeris of \citet{Vogt1981a},
in which $t_0$ represents the time of superior conjunction, when the primary is
farthest from the observer.

The next task was to subdivide the data into individual data sets comprised
of pairs of $B$ and $V$ light curves suitable for inversion. Each light curve
needed to contain enough data points so as to provided good phase coverage.
Ideally this would be achieved using data acquired during a single rotation of
the star, since this would minimize the chance that the spot configuration had
evolved significantly during the time interval spanned by the data. However, in
practice this was not feasible, because the star's rotation period is too long
to allow for continuous monitoring during a single rotation, and because the
telescope was not dedicated solely to this study.
The desire for good phase coverage is thus in opposition to the desire to
minimize the number of stellar rotations covered in any one data set, so
some compromises were necessary. If data 
for different revolutions showed a systematic shift in the magnitudes, it was
clear that the stellar surface features had evolved by an unacceptable amount
during the interval in question; otherwise, data from additional rotations could
be included if needed so as to improve the phase coverage. Several groups of
observations were discarded because they were temporally isolated by many
rotation periods from the observations nearest them in time and contained an
insufficient number of observations to produce good phase coverage. In the end,
68 pairs of $B$ and $V$ light curves were created that were subsequently
inverted to produce the results reported in this study. The mean number of
observations per $B$ light curve was 22.3, the median was 21.5 and the
standard deviation was 6.3, while for the $V$ light curves the mean was 22.1,
the median was 21 and the standard deviation was 5.9. 

The final task before inverting the light curves was to convert the
differential magnitudes to relative intensities. The formula used for both the
$B$ and the $V$ passbands was simply
\begin{equation}
I = 10^{-0.4(m-m_0)},
\end{equation}
where $m$ was the differential magnitude for the observation in question, and
the reference magnitude $m_0$ was the smallest value of $m$ in the entire data set
for the given filter. An intensity of $I \equiv 1$ was thus assigned to this observation.
For the $B$ filter, $m_0 = 1.843$, while for the $V$ filter, $m_0 = 1.522$.
No attempt was made to calibrate the $V$
intensities relative to the $B$ intensities in an absolute sense, because the LI algorithm neither
requires nor would make use of this information, as mentioned in \S
\ref{sec:LI_Algorithm}.

\citet{Berdyugina1998a} obtained $\alpha = 60^\circ \pm 10^\circ$ for the inclination
of the rotation axis. We performed inversions assuming their nominal value
of $\alpha = 60^\circ$. We also performed inversions for $\alpha = 45^\circ$ as well, 
considering this to be prudent given the considerable uncertainty in the inclination.
As will be seen from the discussion of the individual data sets below, the results
for both assumed inclinations were generally consistent, increasing our confidence in their
validity. While one might argue that we also should have performed inversions for
assumed inclinations greater than $60^\circ$, say $70^\circ$ or $75^\circ$, simulations
like those reported in \citet{HarmonCrews2000} show that the method works poorly in such
circumstances. We thus chose not to do so.

For the spot and photosphere temperatures, we used $T_\mathrm{spot} = 3500$~K
since this is compatible with the estimates by other authors mentioned in
\S \ref{sec:Introduction},
and $T_\mathrm{phot} = 4600$~K based on the work of \citet{Berdyugina1998a}. 
The estimates $r_n$
of $J_{n;\mathrm{phot}}/J_{1;\mathrm{phot}}$ and $s_n$ of
$J_{n;\mathrm{spot}}/J_{n;\mathrm{phot}}$ appearing in the scaling given by
equation (\ref{eq:Scaling}) were obtained by evaluating the
Planck function describing blackbody radiation at the filter effective
wavelengths $\lambda_{B,\mathrm{eff}} = 440$~nm and $\lambda_{V,\mathrm{eff}} =
550$~nm.

The photosphere temperature $T_\mathrm{phot}$ and surface acceleration due to
gravity $g$ are the input parameters used by \citet{VanHamme1993} to calculate
limb-darkening coefficients based on the ATLAS stellar atmosphere models of
\citet{Kurucz1991}. We used 
$\log g = 3.0$, which is appropriate for a K2 subgiant \citep{Gray1992}, and is
the value in Van Hamme's tables which is closest to the result $\log g = 3.2$ of
\citet{Berdyugina1998a}. Van Hamme
gives coefficients in steps of 250~K for temperatures in the range $3500\mbox{
K} < T_\mathrm{phot} < 10000\mbox{ K}$, so there is no entry for our value of
$T_\mathrm{phot} = 4600$~K. We simply substituted the values corresponding to
$T_\mathrm{phot} = 4500$~K, the nearest listed temperature to ours, since we do
not know $T_\mathrm{phot}$ or $g$ accurately enough to justify interpolating.
From Table 2 in Van Hamme's paper, we find for the $B$ filter that
\begin{equation}
L_B(\mu) = 1 - \epsilon(1 - \mu) - \delta \mu \ln\mu,
\end{equation}
with $\epsilon = 0.852$ and $\delta = -0.158$, where $\mu$ is the cosine of the
angle between the outward normal to the surface and the observer's line of
sight. For the $V$ filter 
\begin{equation}
L_V(\mu) = 1 - \epsilon(1 - \mu) - \delta(1 - \sqrt{\mu}),
\end{equation}
with $\epsilon = 0.780$ and $\delta = 0.039$. Given the latitude and longitude
of patch $(i,j)$ and the angle of inclination $\alpha$, it is straightforward
to compute $\mu$ for the center of the patch at any given observation time
$t_{nk}$ and thus to obtain $L_{nk;ij}$ in equation (\ref{eq:Ink}).

The next step was to obtain estimates of the noise variances $\sigma_B^2$ and
$\sigma_V^2$ for each of the 68 light curves for each filter. It is well known
that in practice Twomey's criterion of choosing the tradeoff parameter
$\lambda$ so that the variance between the data and the reconstruction is equal
to the noise variance leads to over-smoothing \citep{Turchin1967}. In our
case this would lead to a loss in resolution of the reconstructed surfaces. However,
the technique we use to determine $\lambda$ avoids this problem, because we
obtain an ``effective noise level'' rather than using an estimate of the noise
variance based on the scatter in the comparison star magnitudes. \citet{HarmonCrews2000} 
describes simulations in which an artificial star is used to generate light curves to
which random noise of a known variance $\tilde{\sigma}_n^2$ is added. This
allowed the effects of using an underestimate $\sigma_n^2 < \tilde{\sigma}_n^2$ 
of the noise variance in the goodness-of-fit function 
$G(\mathbf{\hat{J}},\mathbf{I})$ to be
determined. It was found that for a given light curve, when the ratio
$\sigma_n^2/\tilde{\sigma}_n^2$ falls below a certain value (typically between
0.90 and 0.98), the solution ``falls apart'' in that it starts to show very
obvious noise
artifacts. The transition to this behavior is quite sharp in that it takes place over a
narrow range of values of this ratio. This gives a practical means to determine
how low $\sigma_n^2$ can be pushed while still yielding acceptable solutions.
Thus, we can avoid the over-smoothing associated with the Twomey criterion
by performing inversions for a range of values of $\sigma_n^2$ and then choosing
the lowest value which leads to a solution free of obvious noise artifacts.
This value is what we call the ``effective noise level'' for the light curve.
Because the transition is not perfectly sharp, the precise choice of
$\sigma_n^2$ is to some extent a judgment call, so we were conservative in our
choices.

The first round of inversions was thus a series of single-filter inversions of
all 68 $B$ and 68 $V$ light curves so as to assign an effective noise to each
one. The next round was to take each pair of contemporaneous $B$ and $V$ light 
curves and invert the pair in combination so as to
produce the finished surface map for that pair. A complication is that when
using LI to simultaneously invert light curves
obtained through several filters, the effective noise levels as determined
from single-filter inversions sometimes (but not always) lead to under-smoothed
surface images corrupted by noise artifacts. Thus, we used the
effective noise levels obtained from the initial round of single-filter
inversions as starting points, and ran a series of inversions for each pair of
light curves using nearby values of $\sigma_B$ and $\sigma_V$. The resulting
images were inspected to determine the lowest values of $\sigma_B$ and
$\sigma_V$ that did not result in obvious noise artifacts; again we were
conservative in our judgments.

In general, when simultaneously inverting multiple light curves, 
the deviation for a given filter between the data and reconstructed light curves
differs somewhat from the effective noise. This arises because the convergence criterion
for the LI algorithm is based on the \emph{overall} deviation between the data and 
reconstructed light curves through the various filters taken together as a whole, 
rather than on the deviations for the
individual filters. This is necessary because the scaling given by equation (\ref{eq:Scaling}),
which defines the patch intensities assigned to the secondary filter(s) in terms of their
values as seen through the primary filter, makes it impossible to independently tweak how well
the individual reconstructed light curves match the data light curves. 

\section{\label{sec:Results} Results}

Each of the 68 data sets consisting of paired $B$ and $V$ light curves was inverted according 
to the procedure outlined in \S \ref{sec:LI_Algorithm}. Upon 
careful examination of the resulting images, the data sets were divided into
21 groups. Each group covers a span of time during which we can see evolution 
of a particular set of spots, except for three cases in which a group consists of 
a single data set because it is temporally isolated from the data sets immediately before and 
after it by long gaps in the data.
In six of these groups we saw good evidence for
differential rotation, and as looking for such evidence was our primary goal in 
this work, we present in detail only the results for these six groups 
here, arranged in chronological order. The time spans corresponding to these six groups
are indicated via horizontal bars in the plots of Figure \ref{fig:CompleteDataSet}. 
The remainder of our images are presented in Figures 17--33 in the online version of the
Journal.

Table \ref{tab:DataSets} summarizes the properties of each of the data sets and groups. 
The data sets are numbered in chronological order in the first column, with the group to which 
each set was assigned indicated in the second column. Boldface entries in the first two columns
denote the groups for which we present evidence of differential rotation. The starting and ending
heliocentric Modified Julian Dates and corresponding UTC calendar dates for the 
$B$ and $V$ light curves of each data set are given in the fourth through seventh columns; 
a blank entry for a $V$ light curve
indicates that the value is the same as for the $B$ light curve from the same set. The number of
data points in each data set is shown in the $N_\mathrm{obs}$ column. The number of rotation
periods covered by each data set is listed in the ``\# Per.'' column. The last two columns list
the ``effective noise'' in magnitudes for each inversion, as described in section 
\S \ref{sec:DataAnalysis}, for the assumed rotation axis inclinations of $\alpha = 45^\circ$
and $\alpha = 60^\circ$. The values in the table are $10^4$ times larger than the actual values, e.g.,
an entry of ``154'' means that the effective noise was 0.0154~mag.

\begin{deluxetable}{rrrrrrrrrrr}
\tabletypesize{\scriptsize}
\tablewidth{0pt}
\tablecolumns{11}
\tablecaption{\label{tab:DataSets} Light Curve Data Sets and Groups}
\tablehead{
\colhead{Set}	&	\colhead{Grp.}	&	\colhead{Filt.}	&	\colhead{Start MJD}	&	\colhead{End MJD}	&	\colhead{Start Date}	&	\colhead{End Date}	&	\colhead{$N_\mathrm{obs}$}	&	\colhead{\# Per.}	&	\colhead{$\sigma_\mathrm{eff,45}$}	&	
\colhead{$\sigma_\mathrm{eff,60}$}}
\startdata														
1	&	1	&	$B$	&	47115.8086	&	47141.7351	&	1987 Nov 16	&	1987 Dec 12	&	15	&	3.86	&	154	&	154	\\									
	&		&	$V$	&		&		&		&		&	17	&		&	194	&	194	\\									
2	&	1	&	$B$	&	47171.6466	&	47198.5822	&	1988 Jan 11	&	1988 Feb 07	&	10	&	4.01	&	138	&	132	\\									
	&		&	$V$	&		&	47199.5824	&		&	1988 Feb 08	&	12	&	4.15	&	226	&	216	\\									
\textbf{3}	&	\textbf{2}	&	$B$	&	47417.7681	&	47433.8609	&	1988 Sep 13	&	1988 Sep 29	&	20	&	2.39	&	268	&	258	\\									
	&		&	$V$	&	47415.8437	&		&	1988 Sep 11	&		&	23	&	2.68	&	326	&	326	\\									
\textbf{4}	&	\textbf{2}	&	$B$	&	47434.7230	&	47469.8305	&	1988 Sep 30	&	1988 Nov 04	&	32	&	5.22	&	164	&	158	\\									
	&		&	$V$	&		&		&		&		&	30	&		&	208	&	200	\\
\textbf{5}	&	\textbf{2}	&	$B$	&	47470.6837	&	47525.6801	&	1988 Nov 05	&	1988 Dec 30	&	31	&	8.18	&	174	&	174	\\									
	&		&	$V$	&		&		&		&		&	28	&		&	294	&	286	\\							
\textbf{6}	&	\textbf{2}	&	$B$	&	47526.5929	&	47556.5813	&	1988 Dec 31	&	1989 Jan 30	&	12	&	4.46	&	144	&	144	\\									
	&		&	$V$	&		&	47549.5824	&		&	1989 Jan 23	&	13	&	3.42	&	146	&	146	\\		
						
\textbf{7}	&	\textbf{3}	&	$B$	&	47779.7480	&	47818.8716	&	1989 Sep 10	&	1989 Oct 19	&	35	&	5.82	&	304	&	296	\\									
	&		&	$V$	&		&		&		&		&	33	&		&	298	&	298	\\									
\textbf{8}	&	\textbf{3}	&	$B$	&	47824.6882	&	47850.7976	&	1989 Oct 25	&	1989 Nov 20	&	30	&	3.88	&	392	&	390	\\									
	&		&	$V$	&		&		&		&		&	31	&		&	410	&	402	\\									
\textbf{9}	&	\textbf{3}	&	$B$	&	47853.6658	&	47921.5825	&	1989 Nov 23	&	1990 Jan 30	&	38	&	10.10	&	238	&	214	\\								
	&		&	$V$	&		&	47926.5752	&		&	1990 Feb 04	&	33	&	10.84	&	238	&	206	\\									
10	&	4	&	$B$	&	48183.8086	&	48236.6947	&	1990 Oct 19	&	1990 Dec 11	&	25	&	7.86	&	278	&	278	\\									
	&		&	$V$	&		&	48235.6694	&		&	1990 Dec 10	&	25	&	7.71	&	182	&	184	\\									
11	&	5	&	$B$	&	48397.9614	&	48437.9355	&	1991 May 21	&	1991 Jun 30	&	21	&	5.94	&	124	&	118	\\									
	&		&	$V$	&	48394.9734	&		&	1991 May 18	&		&	27	&	6.39	&	94	&	94	\\									
12	&	6	&	$B$	&	48872.9343	&	48898.8417	&	1992 Sep 07	&	1992 Oct 03	&	17	&	3.85	&	82	&	80	\\									
	&		&	$V$	&		&		&		&		&	17	&		&	56	&	64	\\									
13	&	6	&	$B$	&	48905.7999	&	48945.7709	&	1992 Oct 10	&	1992 Nov 19	&	28	&	5.94	&	200	&	196	\\									
	&		&	$V$	&		&		&		&		&	28	&		&	196	&	198	\\									
14	&	6	&	$B$	&	48951.7163	&	49022.5799	&	1992 Nov 25	&	1993 Feb 04	&	20	&	10.54	&	112	&	106	\\									
	&		&	$V$	&		&		&		&		&	20	&		&	106	&	106	\\									
15	&	7	&	$B$	&	49135.9488	&	49166.9202	&	1993 May 28	&	1993 Jun 28	&	23	&	4.61	&	124	&	128	\\									
	&		&	$V$	&		&		&		&		&	23	&		&	122	&	118	\\									
16	&	7	&	$B$	&	49235.9074	&	49253.7684	&	1993 Sep 05	&	1993 Sep 23	&	15	&	2.66	&	46	&	40	\\									
	&		&	$V$	&		&		&		&		&	15	&		&	44	&	46	\\									
17	&	7	&	$B$	&	49275.8128	&	49340.6260	&	1993 Oct 15	&	1993 Dec 19	&	25	&	9.64	&	144	&	140	\\									
	&		&	$V$	&	49282.7438	&		&	1993 Oct 22	&		&	26	&	8.61	&	132	&	134	\\									
18	&	7	&	$B$	&	49347.6242	&	49384.5901	&	1993 Dec 26	&	1994 Feb 01	&	17	&	5.50	&	94	&	106	\\									
	&		&	$V$	&	49345.6833	&		&	1993 Dec 24	&		&	18	&	5.79	&	104	&	116	\\									
19	&	8	&	$B$	&	49638.7416	&	49671.8291	&	1994 Oct 13	&	1994 Nov 15	&	22	&	4.92	&	106	&	110	\\									
	&		&	$V$	&		&		&		&		&	22	&		&	104	&	108	\\									
20	&	8	&	$B$	&	49674.7171	&	49706.6550	&	1994 Nov 18	&	1994 Dec 20	&	17	&	4.75	&	56	&	56	\\									
	&		&	$V$	&		&		&		&		&	18	&		&	76	&	76	\\									
21	&	8	&	$B$	&	49724.6199	&	49757.5937	&	1995 Jan 07	&	1995 Feb 09	&	14	&	4.90	&	70	&	76	\\									
	&		&	$V$	&		&		&		&		&	14	&		&	86	&	96	\\									
22	&	9	&	$B$	&	49873.9567	&	49909.8632	&	1995 Jun 05	&	1995 Jul 11	&	26	&	5.34	&	120	&	116	\\									
	&		&	$V$	&		&		&		&		&	25	&		&	110	&	114	\\									
23	&	10	&	$B$	&	49982.8845	&	50001.9313	&	1995 Sep 22	&	1995 Oct 11	&	27	&	2.83	&	88	&	96	\\									
	&		&	$V$	&		&		&		&		&	28	&		&	54	&	58	\\									
24	&	10	&	$B$	&	50002.7800	&	50033.7425	&	1995 Oct 12	&	1995 Nov 12	&	32	&	4.60	&	88	&	86	\\									
	&		&	$V$	&		&		&		&		&	32	&		&	82	&	82	\\									
25	&	10	&	$B$	&	50037.8137	&	50062.6899	&	1995 Nov 16	&	1995 Dec 11	&	27	&	3.70	&	146	&	140	\\									
	&		&	$V$	&		&		&		&		&	26	&		&	120	&	116	\\									
26	&	10	&	$B$	&	50066.7615	&	50123.5946	&	1995 Dec 15	&	1996 Feb 10	&	30	&	8.45	&	162	&	164	\\									
	&		&	$V$	&		&	50129.5853	&		&	1996 Feb 16	&	30	&	9.34	&	132	&	132	\\									
\textbf{27}	&	\textbf{11}	&	$B$	&	50391.8348	&	50435.6313	&	1996 Nov 04	&	1996 Dec 18	&	26	&	6.51	&	120	&	124	\\									
	&		&	$V$	&		&		&		&		&	25	&		&	114	&	104	\\									
\textbf{28}	&	\textbf{11}	&	$B$	&	50436.6379	&	50490.5894	&	1996 Dec 19	&	1997 Feb 11	&	26	&	8.02	&	208	&	204	\\									
	&		&	$V$	&		&	50494.5846	&		&	1997 Feb 15	&	25	&	8.62	&	152	&	140	\\									
\textbf{29}	&	\textbf{11}	&	$B$	&	50590.9563	&	50642.8849	&	1997 May 22	&	1997 Jul 13	&	25	&	7.72	&	116	&	104	\\									
	&		&	$V$	&		&		&		&		&	28	&		&	108	&	100	\\									
\textbf{30}	&	\textbf{11}	&	$B$	&	50714.8051	&	50755.7973	&	1997 Sep 23	&	1997 Nov 03	&	35	&	6.10	&	120	&	120	\\									
	&		&	$V$	&		&		&		&		&	35	&		&	126	&	108	\\									
\textbf{31}	&	\textbf{11}	&	$B$	&	50756.7863	&	50795.6926	&	1997 Nov 04	&	1997 Dec 13	&	23	&	5.79	&	72	&	80	\\									
	&		&	$V$	&		&		&		&		&	21	&		&	70	&	78	\\									
\textbf{32}	&	\textbf{11}	&	$B$	&	50797.6863	&	50837.6257	&	1997 Dec 15	&	1998 Jan 24	&	20	&	5.94	&	124	&	114	\\									
	&		&	$V$	&		&		&		&		&	19	&		&	118	&	104	\\									
\textbf{33}	&	\textbf{11}	&	$B$	&	50838.6203	&	50856.5946	&	1998 Jan 25	&	1998 Feb 12	&	13	&	2.67	&	172	&	180	\\									
	&		&	$V$	&		&		&		&		&	13	&		&	120	&	128	\\									
\textbf{34}	&	\textbf{12}	&	$B$	&	51085.9791	&	51115.7864	&	1998 Sep 29	&	1998 Oct 29	&	25	&	4.43	&	86	&	86	\\									
	&		&	$V$	&	51086.7915	&		&	1998 Sep 30	&		&	25	&	4.31	&	116	&	96	\\									
\textbf{35}	&	\textbf{12}	&	$B$	&	51116.7725	&	51144.7050	&	1998 Oct 30	&	1998 Nov 27	&	28	&	4.15	&	120	&	122	\\									
	&		&	$V$	&		&		&		&		&	27	&		&	112	&	108	\\									
\textbf{36}	&	\textbf{12}	&	$B$	&	51148.7022	&	51182.6674	&	1998 Dec 01	&	1999 Jan 04	&	24	&	5.05	&	76	&	84	\\									
	&		&	$V$	&		&		&		&		&	24	&		&	104	&	114	\\									
\textbf{37}	&	\textbf{12}	&	$B$	&	51183.6640	&	51224.5942	&	1999 Jan 05	&	1999 Feb 15	&	24	&	6.09	&	172	&	174	\\									
	&		&	$V$	&		&		&		&		&	23	&		&	162	&	158	\\									
\textbf{38}	&	\textbf{13}	&	$B$	&	51429.8915	&	51475.8034	&	1999 Sep 08	&	1999 Oct 24	&	29	&	6.83	&	146	&	146	\\									
	&		&	$V$	&		&		&		&		&	28	&		&	108	&	108	\\									
\textbf{39}	&	\textbf{13}	&	$B$	&	51477.8148	&	51505.7306	&	1999 Oct 26	&	1999 Nov 23	&	27	&	4.15	&	202	&	208	\\									
	&		&	$V$	&		&		&		&		&	24	&		&	118	&	130	\\									
\textbf{40}	&	\textbf{13}	&	$B$	&	51506.7278	&	51535.6724	&	1999 Nov 24	&	1999 Dec 23	&	23	&	4.30	&	94	&	100	\\									
	&		&	$V$	&		&		&		&		&	23	&		&	76	&	72	\\									
\textbf{41}	&	\textbf{13}	&	$B$	&	51537.6574	&	51586.5994	&	1999 Dec 25	&	2000 Feb 12	&	29	&	7.28	&	152	&	138	\\									
	&		&	$V$	&		&		&		&		&	30	&		&	118	&	126	\\									
\textbf{42}	&	\textbf{14}	&	$B$	&	51805.8758	&	51833.8204	&	2000 Sep 18	&	2000 Oct 16	&	17	&	4.16	&	124	&	126	\\									
	&		&	$V$	&		&		&		&		&	18	&		&	154	&	156	\\									
\textbf{43}	&	\textbf{14}	&	$B$	&	51838.8332	&	51879.8006	&	2000 Oct 21	&	2000 Dec 01	&	18	&	6.09	&	80	&	74	\\									
	&		&	$V$	&		&		&		&		&	18	&		&	94	&	84	\\									
\textbf{44}	&	\textbf{14}	&	$B$	&	51884.7888	&	51945.6130	&	2000 Dec 06	&	2001 Feb 05	&	25	&	9.05	&	130	&	136	\\									
	&		&	$V$	&	51886.6803	&	51946.6051	&	2000 Dec 08	&	2001 Feb 06	&	21	&	8.91	&	180	&	176	\\									
45	&	15	&	$B$	&	52178.8580	&	52214.7927	&	2001 Sep 26	&	2001 Nov 01	&	22	&	5.34	&	152	&	152	\\									
	&		&	$V$	&		&		&		&		&	22	&		&	128	&	130	\\									
46	&	15	&	$B$	&	52216.7903	&	52261.6769	&	2001 Nov 03	&	2001 Dec 18	&	17	&	6.68	&	168	&	150	\\									
	&		&	$V$	&		&		&		&		&	15	&		&	120	&	118	\\									
47	&	15	&	$B$	&	52266.7173	&	52313.6093	&	2001 Dec 23	&	2002 Feb 08	&	16	&	6.97	&	124	&	130	\\	
	&		&	$V$	&	52267.6603	&		&	2001 Dec 24	&		&	15	&	6.83	&	154	&	130	\\									
48	&	16	&	$B$	&	52539.8107	&	52595.7358	&	2002 Sep 22	&	2002 Nov 17	&	34	&	8.32	&	136	&	140	\\									
	&		&	$V$	&		&		&		&		&	33	&		&	120	&	118	\\									
49	&	16	&	$B$	&	52597.7373	&	52645.6685	&	2002 Nov 19	&	2003 Jan 06	&	29	&	7.13	&	100	&	102	\\									
	&		&	$V$	&		&		&		&		&	30	&		&	88	&	94	\\									
50	&	16	&	$B$	&	52649.6412	&	52677.5907	&	2003 Jan 10	&	2003 Feb 07	&	22	&	4.16	&	134	&	124	\\									
	&		&	$V$	&		&		&		&		&	21	&		&	80	&	78	\\									
51	&	17	&	$B$	&	52894.9328	&	52914.8841	&	2003 Sep 12	&	2003 Oct 02	&	17	&	2.97	&	96	&	94	\\									
	&		&	$V$	&		&		&		&		&	18	&		&	164	&	188	\\									
52	&	17	&	$B$	&	52915.8656	&	52941.8171	&	2003 Oct 03	&	2003 Oct 29	&	19	&	3.86	&	212	&	206	\\									
	&		&	$V$	&		&	52944.8047	&		&	2003 Nov 01	&	20	&	4.30	&	134	&	132	\\									
53	&	17	&	$B$	&	52947.7834	&	52986.7346	&	2003 Nov 04	&	2003 Dec 13	&	18	&	5.79	&	78	&	62	\\									
	&		&	$V$	&		&		&		&		&	19	&		&	72	&	70	\\									
54	&	17	&	$B$	&	52988.7597	&	53035.6267	&	2003 Dec 15	&	2004 Jan 31	&	21	&	6.97	&	184	&	190	\\									
	&		&	$V$	&		&		&		&		&	20	&		&	118	&	118	\\									
55	&	18	&	$B$	&	53255.8137	&	53284.8259	&	2004 Sep 07	&	2004 Oct 06	&	18	&	4.31	&	104	&	102	\\									
	&		&	$V$	&		&		&		&		&	18	&		&	54	&	62	\\									
56	&	18	&	$B$	&	53285.8297	&	53315.7442	&	2004 Oct 07	&	2004 Nov 06	&	17	&	4.45	&	70	&	70	\\									
	&		&	$V$	&		&		&		&		&	17	&		&	68	&	62	\\									
57	&	18	&	$B$	&	53326.7403	&	53355.6576	&	2004 Nov 17	&	2004 Dec 16	&	17	&	4.30	&	66	&	66	\\									
	&		&	$V$	&		&		&		&		&	18	&		&	64	&	66	\\									
58	&	18	&	$B$	&	53357.6446	&	53405.6231	&	2004 Dec 18	&	2005 Feb 04	&	20	&	7.14	&	114	&	128	\\									
	&		&	$V$	&		&		&		&		&	19	&		&	146	&	156	\\									
59	&	19	&	$B$	&	53521.9609	&	53566.8945	&	2005 May 31	&	2005 Jul 15	&	21	&	6.68	&	136	&	134	\\									
	&		&	$V$	&		&		&		&		&	21	&		&	124	&	124	\\									
60	&	20	&	$B$	&	53627.6919	&	53646.7436	&	2005 Sep 14	&	2005 Oct 03	&	20	&	2.83	&	140	&	142	\\									
	&		&	$V$	&		&		&		&		&	20	&		&	86	&	88	\\									
61	&	20	&	$B$	&	53648.7212	&	53673.7304	&	2005 Oct 05	&	2005 Oct 30	&	16	&	3.72	&	74	&	92	\\									
	&		&	$V$	&		&		&		&		&	16	&		&	42	&	52	\\									
62	&	20	&	$B$	&	53676.7199	&	53704.6848	&	2005 Nov 02	&	2005 Nov 30	&	18	&	4.16	&	116	&	104	\\									
	&		&	$V$	&	53677.7290	&		&	2005 Nov 03	&		&	14	&	4.01	&	98	&	106	\\									
63	&	20	&	$B$	&	53706.6824	&	53734.6434	&	2005 Dec 02	&	2005 Dec 30	&	13	&	4.16	&	80	&	82	\\									
	&		&	$V$	&		&		&		&		&	14	&		&	70	&	72	\\									
64	&	20	&	$B$	&	53742.6353	&	53773.6036	&	2006 Jan 07	&	2006 Feb 07	&	17	&	4.61	&	62	&	54	\\									
	&		&	$V$	&		&		&		&		&	17	&		&	86	&	84	\\									
65	&	21	&	$B$	&	53875.9673	&	53906.8918	&	2006 May 20	&	2006 Jun 20	&	18	&	4.60	&	72	&	76	\\									
	&		&	$V$	&	53873.9700	&	53907.9678	&	2006 May 18	&	2006 Jun 21	&	20	&	5.06	&	90	&	96	\\									
66	&	21	&	$B$	&	53995.8552	&	54031.7949	&	2006 Sep 17	&	2006 Oct 23	&	19	&	5.34	&	102	&	96	\\									
	&		&	$V$	&		&		&		&		&	19	&		&	72	&	60	\\									
67	&	21	&	$B$	&	54040.7623	&	54094.6978	&	2006 Nov 01	&	2006 Dec 25	&	26	&	8.02	&	108	&	106	\\									
	&		&	$V$	&		&		&		&		&	25	&		&	92	&	90	\\									
68	&	21	&	$B$	&	54103.6923	&	54136.6221	&	2007 Jan 03	&	2007 Feb 05	&	13	&	4.90	&	110	&	102	\\									
	&		&	$V$	&		&	54134.6225	&		&	2007 Feb 03	&	12	&	4.60	&	104	&	88								
\enddata
\end{deluxetable}

\citet{Berdyugina1998b,Berdyugina1999} produced Doppler images of II Peg for 1992--98 
using what they call the ``Occamian'' approach. We comment here on how their images
compare qualitatively to our photometric images obtained at nearly the same time.

First, it should be noted that there is a consistent tendency for corresponding spots 
to appear at higher latitudes in their Doppler images in comparison to our photometric 
images. This is likely due to
our having only two filter passbands available, $B$ and $V$. As shown in \citet{HarmonCrews2000},
having data through four passbands available (e.g., \emph{BVRI}) significantly 
enhances the latitude resolution of the photometric inversions compared to when 
only two are used. Furthermore, a spectral line profile is effectively
a one-dimensional image of the stellar surface \citep{VogtPenrod1983}, while the individual data
in a light curve are zero-dimensional images in that only the integrated
brightness of the surface is indicated by them. Thus, Doppler imaging has intrinsically higher
resolution than does photometric imaging, so it is reasonable to presume that the latitudes
obtained from the Doppler images are more reliable, and thus 
we caution the reader that the latitudes
of the spots on our photometric images should not be accepted at face value. However, the 
extensive simulations reported in \citet{HarmonCrews2000} show that \emph{relative} spot latitudes
can be reliably determined from LI: when two spots at different latitudes are present, the
lower-latitude spot is generally rendered as such, even though the individual spot latitudes
may not be accurately rendered, particularly when only two-filter data are available.
Thus, we should be able to detect differential rotation in our photometric images, though
we cannot go as far as to reliably determine the rotational period as a function of
latitude on the stellar surface.

The 1992 August image of \citet{Berdyugina1998b} compares favorably to our Data Set 12 
(1992 September 7 -- October 3) photometric image, though 
the spot latitudes on the Doppler images are considerably higher. In both images, a large 
spot has an appendage projecting toward lower longitudes. Their 1993 December image shows 
two distinct high-latitude spots almost $180^\circ$ apart,
with one moderately larger than the other. Our Set 18 (1993 December 24 -- 1994 February 1) 
image shows 
what appear to be two widely-separated spots with a ``bridge'' connecting them. The bridge may
be simply an artifact of the inversion, as such bridges are seen in simulations
in which two spots close to one another are present on the artificial stellar surface.
Their 1994 November and 1995 January images show a pair of spots at higher latitudes than our 
Group 8 (1994 October 13 -- 1995 February 9) images. Our images show a pair of spots connected 
by a bridge.
Their 1995 July image and our Set 22 (1995 June 5 -- July 11) image both show a 
high-latitude feature which extends across a wide range in longitude. 
Their 1995 October image and our Group 10 (1995 September 22 -- 1996 February 16) 
images show a pair of 
isolated spots well-separated in longitude, with one larger than the other.

Their Doppler image for 1996 October shows II Peg shortly before one of the time intervals
over which we see good evidence for differential rotation, that of Group 11 
(1996 November 4 -- 1998 February 12), and their images for 1997 June, August, and December
show the star during the interval covered by our images. Our results for this interval are
described in \S \ref{sec:Group11}
below. Similarly, their Doppler images for 1998 October and November represent the
star during another interval over which we saw evidence of differential rotation, 
that of Group 12 (1998 September 29 -- 1999 February 15),
which we discuss in \S \ref{sec:Group12}.
We defer the comparison of their Doppler and our photometric images for these intervals
to those sections.

\subsection{\label{sec:Group2} Group 2: MJD 47417.7681 -- 47556.5813}

\begin{figure}[!htbp]
\centering
\epsscale{0.8}
\plottwo{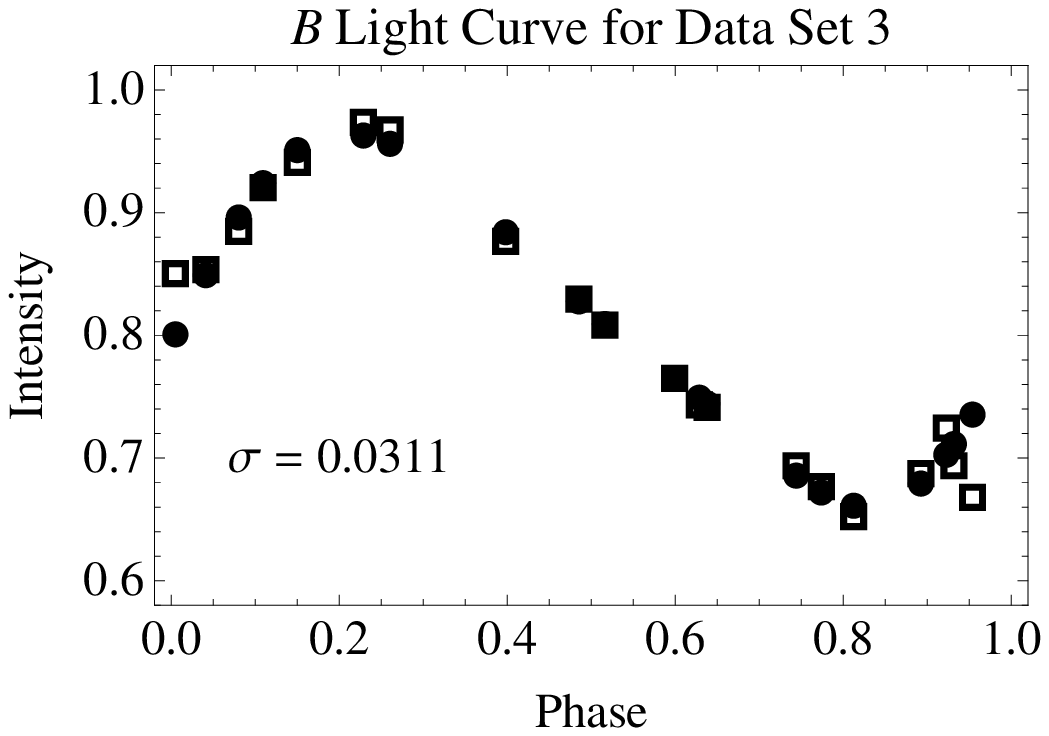}{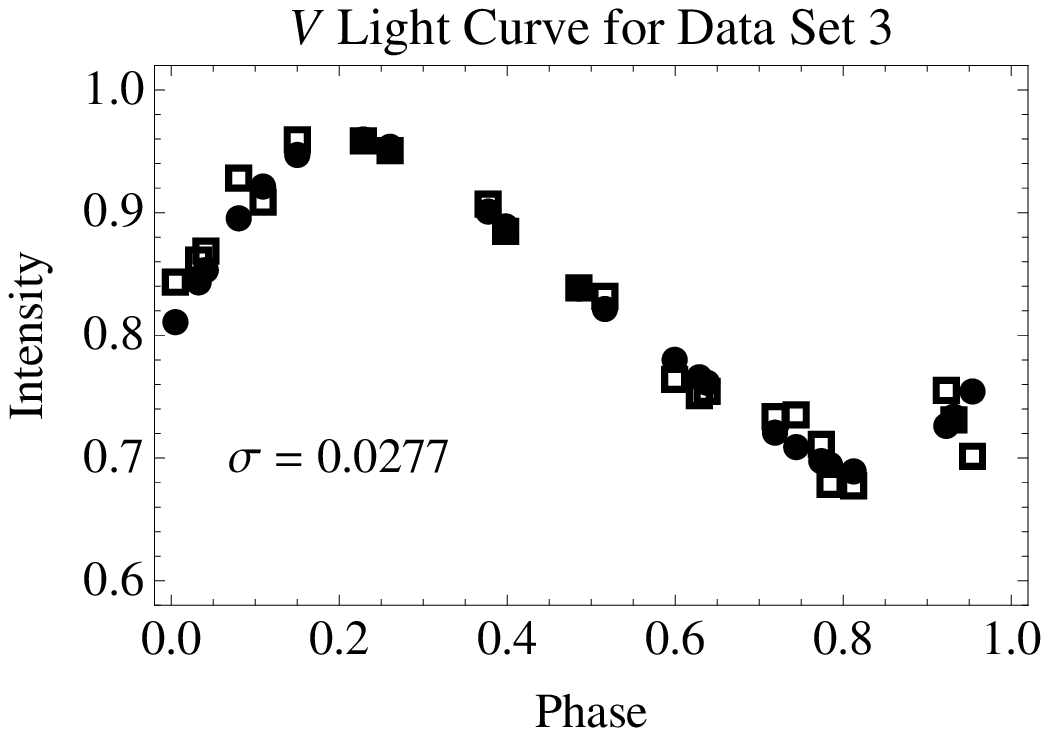}
\vspace{6pt}
\plottwo{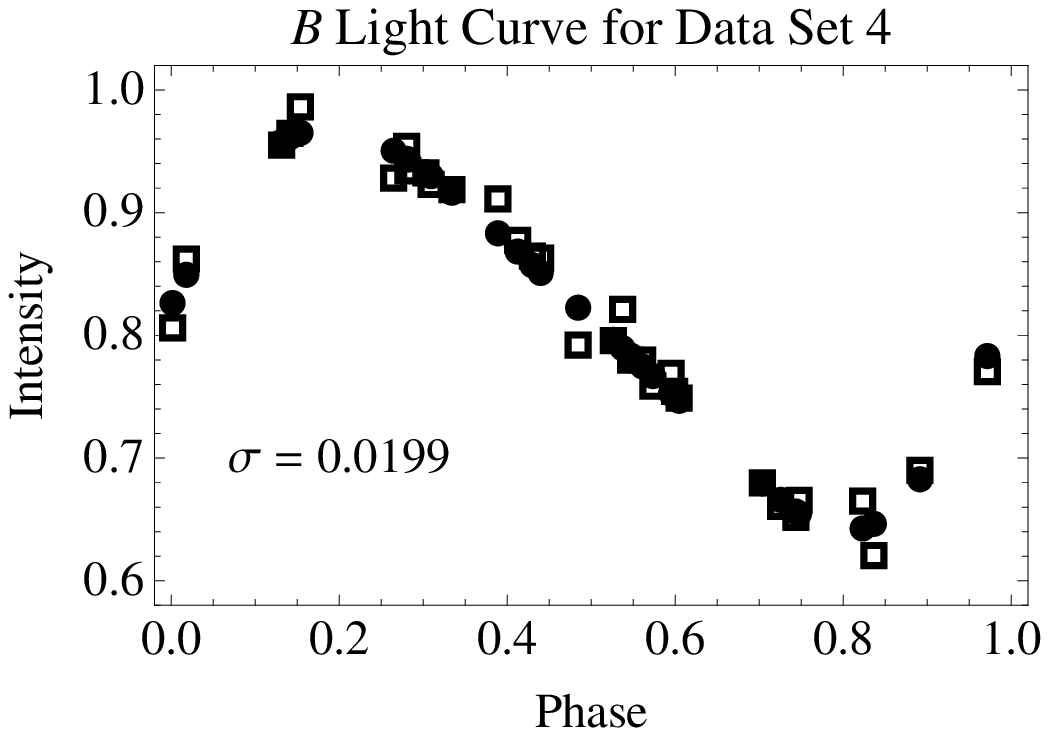}{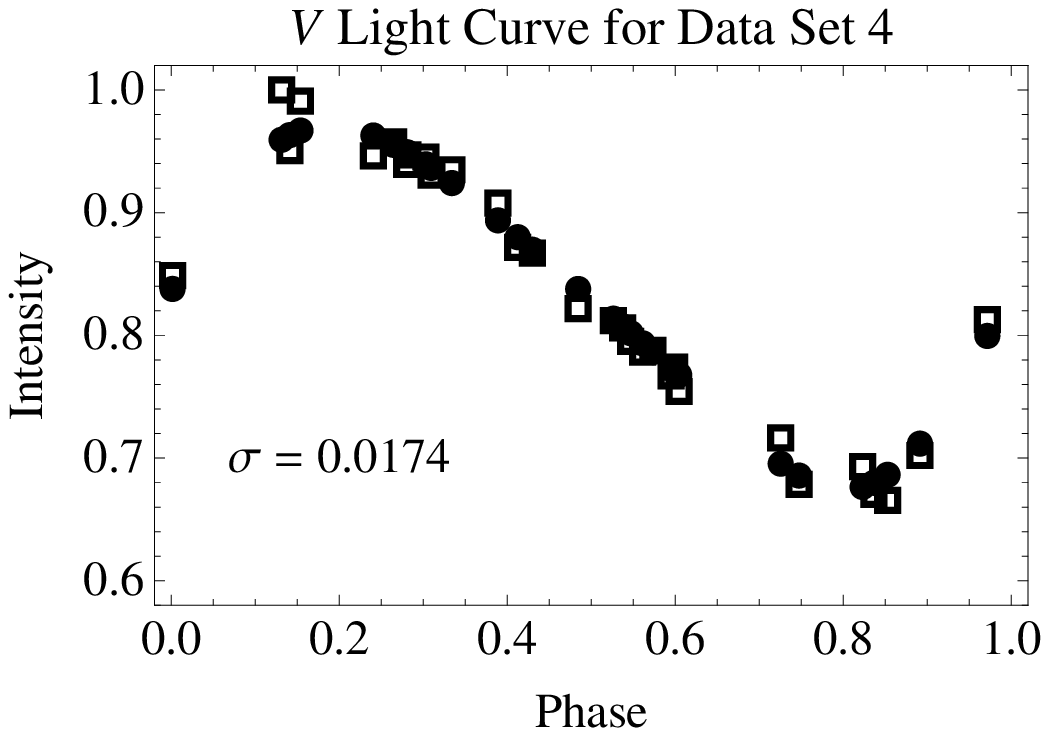}
\vspace{6pt}
\plottwo{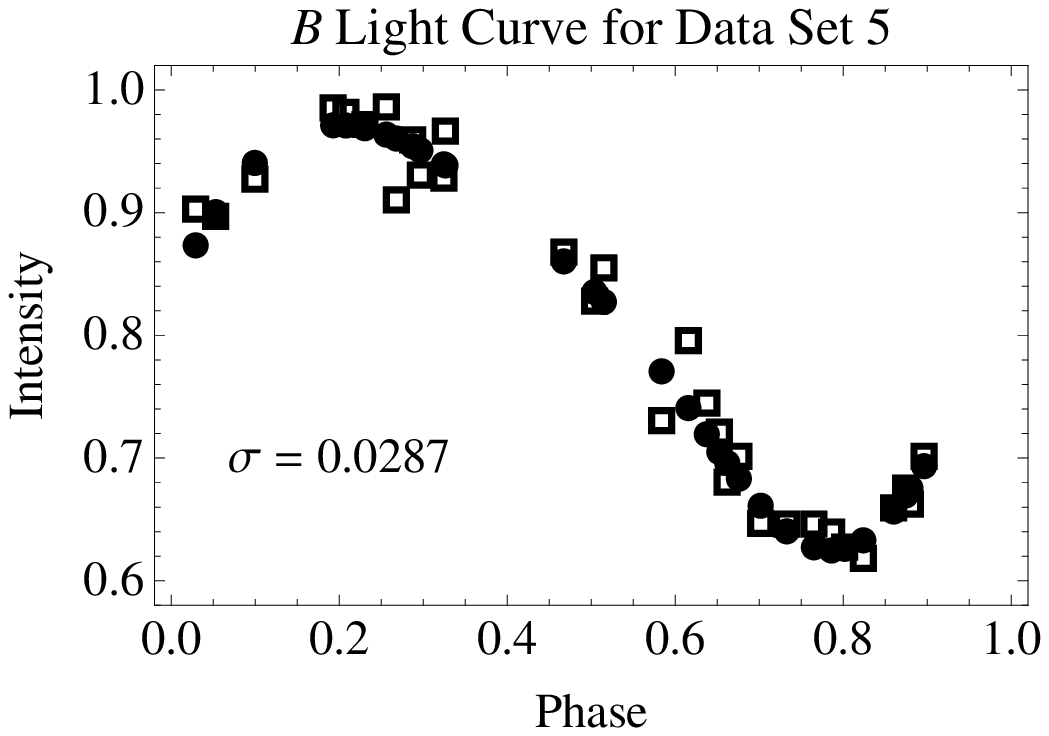}{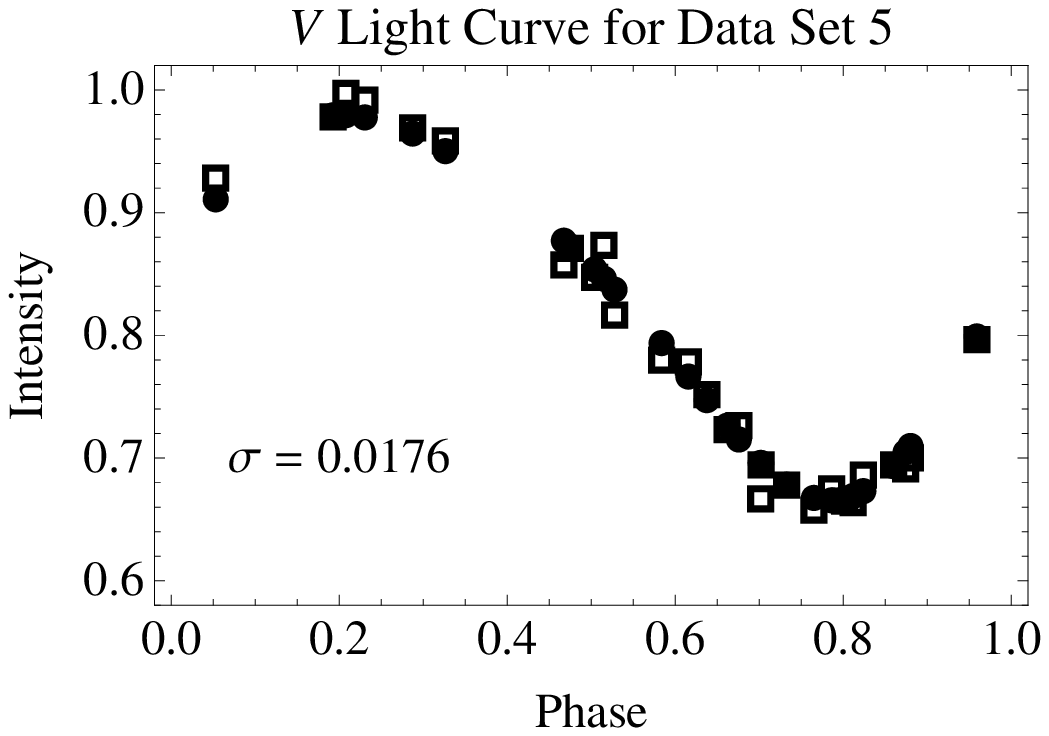}
\vspace{6pt}
\plottwo{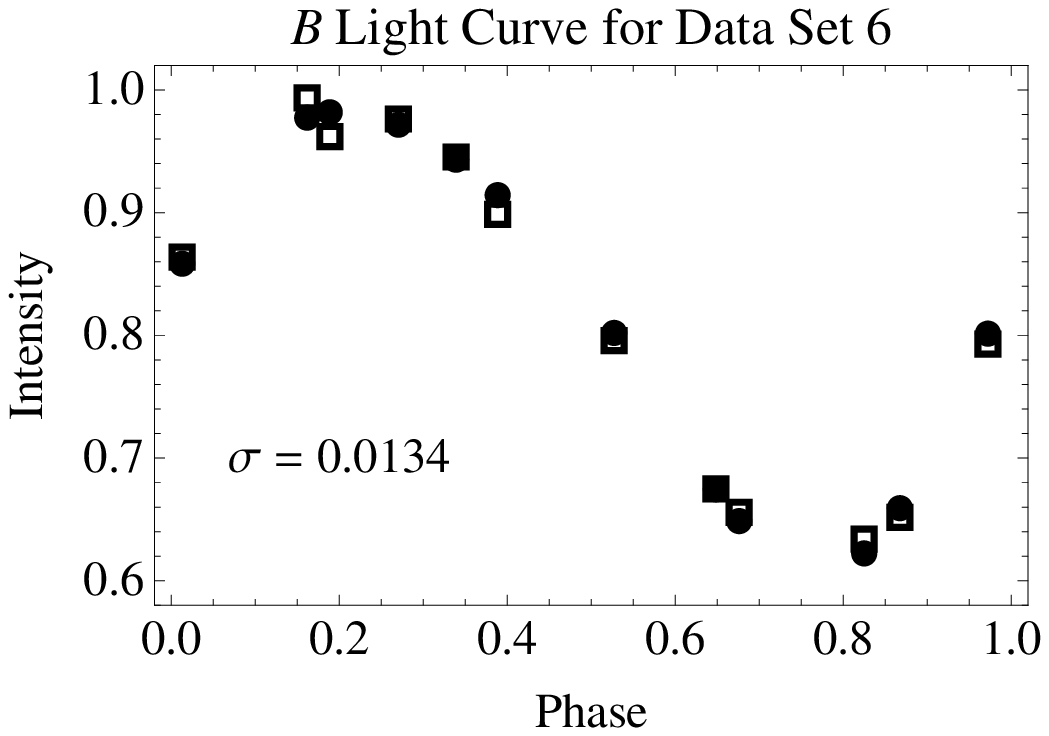}{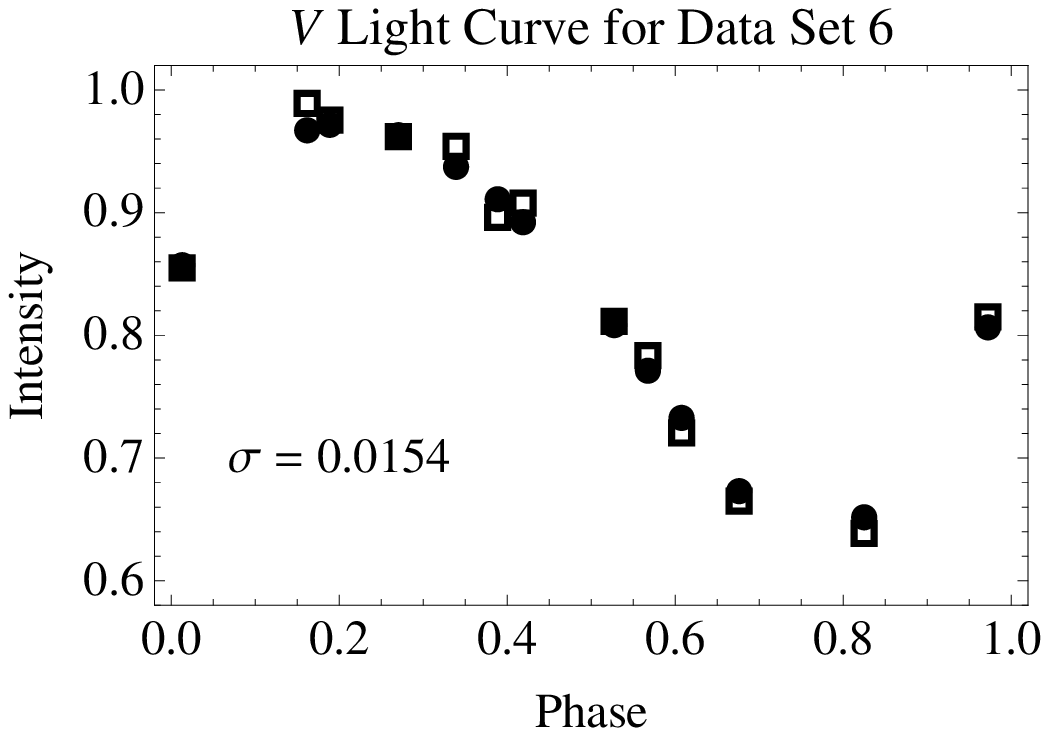}
\epsscale{1.0}
\caption{\label{fig:Group2Plots} $B$- and $V$-filter normalized
intensities (squares) and reconstructed
intensities (filled circles) for the four data sets assigned to Group 2. The assumed
inclination angle between the stellar rotation axis and the line of sight is
$\alpha = 45^\circ$. The RMS deviation $\sigma$ between the
data and reconstructed intensities is indicated on each plot, expressed in magnitudes. 
The corresponding surface 
images are shown in the top row of Figure \ref{fig:Group2}.}
\end{figure}

Figure \ref{fig:Group2Plots} shows plots of the light curves for the four data sets comprising
Group 2, which span the interval from heliocentric MJD 47417.7681 -- 47556.5813 
(1988 September 11 -- 1989 January 30). Also shown in Figure \ref{fig:Group2Plots} are
the reconstructed light curves, i.e., the light curves computed from the stellar surface maps, 
for the cases in which the assumed inclination of the rotation axis to the line of sight
is $\alpha = 45^\circ$. The intensities
have been normalized such that the brightest datum over the entire data set considered in this
study for the given filter is set equal to 1. The quantity $\sigma$ on each plot represents the
rms deviation expressed in magnitudes between the data light curve and the reconstructed light curve.

\begin{figure}[!htbp]
\centering
\plotone{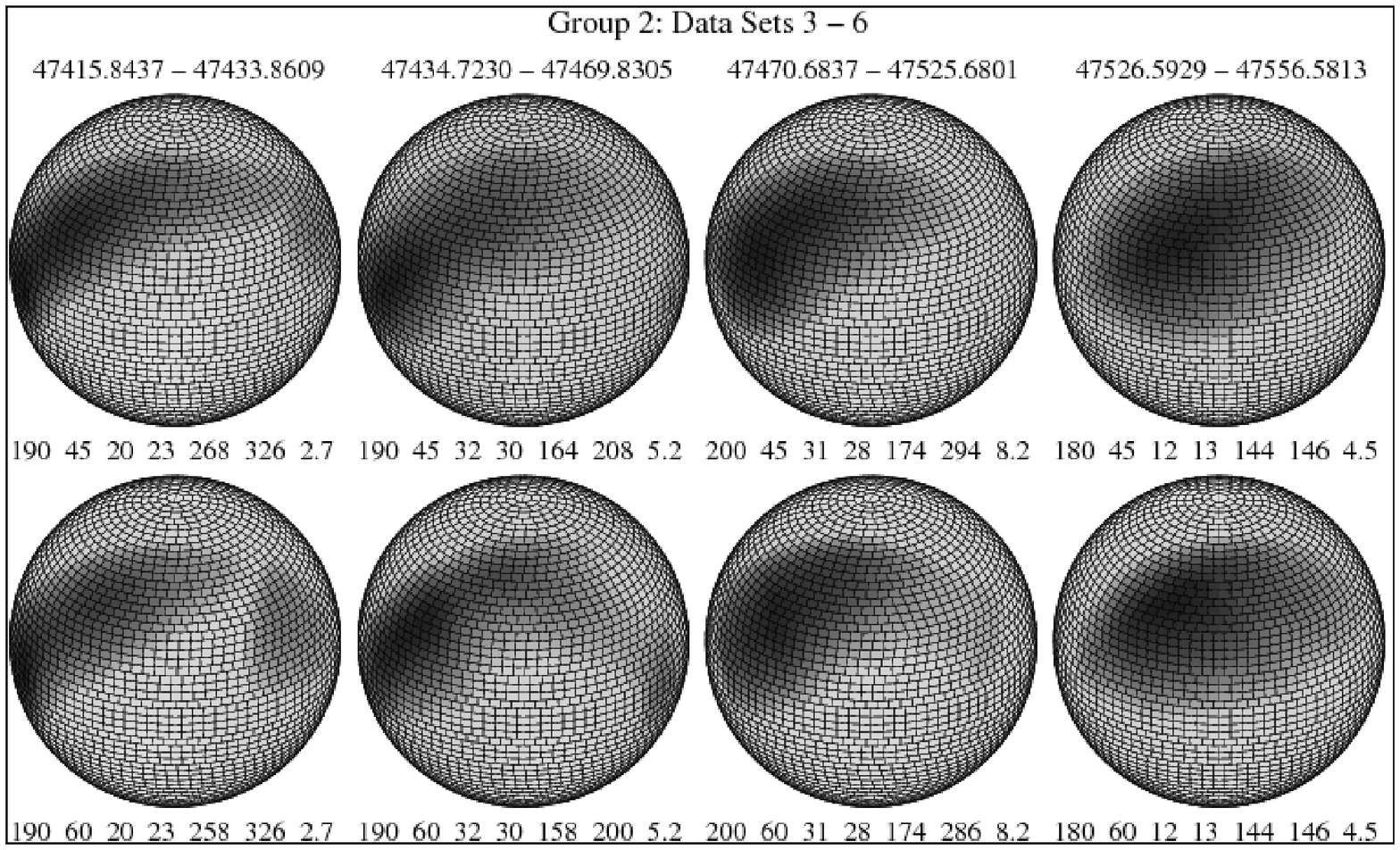}
\caption{\label{fig:Group2} Inversions of the four data sets assigned to Group 2,
acquired between heliocentric MJD 47417.7681 -- 47556.5813 (1988 September 11 -- 1989 January 30) 
The top row is for an assumed inclination
angle between the rotation axis and the line of sight of $\alpha = 45^\circ$,
while the bottom row is for $\alpha = 60^\circ$. The heliocentric
MJD range spanned by each data set is given at the top of each column, so that, for
example, the top image in the third column represents the inversion for $\alpha
= 45^\circ$ of Set 5, acquired from MJD 47470.6837 -- 47525.6801,
while the bottom image in the same column is for $\alpha = 60^\circ$ for the
same data set. The meanings of the numbers below the images are discussed in
\S \ref{sec:Group2}. Inversions shown in subsequent
figures are similarly organized unless otherwise noted.}
\end{figure}

Figure \ref{fig:Group2} shows the reconstructed
surfaces obtained from these light curves via the LI procedure, with the 
results for an assumed axial inclination of $\alpha = 45^\circ$ shown in the top row, 
and for $\alpha = 60^\circ$ shown in 
the bottom row. Each column of two images consists of the pair of $\alpha = 
45^\circ$ and $\alpha = 60^\circ$ reconstructions for the date ranges indicated
at the top of the column. Note that if we let 
$\phi \rightarrow -\phi$ in the specific intensity distribution $J(\theta,\phi)$, 
where $\theta$ and $\phi$ are stellar
latitude and longitude, and also reverse the direction of stellar rotation, the
resulting rotational light curve is unchanged. Thus, based on photometry alone
we cannot distinguish between a given brightness distribution and its mirror
image. For definiteness, in generating our images we have adopted the convention
that the star rotates counterclockwise as viewed from above the visible pole,
so that spots would be carried across our view from left to right with the
visible pole at the top of the image.

We do not show here the other 64 light curves which we inverted. Figure 
\ref{fig:Group2Plots} is included so that the reader may get a sense
of the comparison between the data and reconstructed light curves for a typical
example. However, information about the characteristics of the source light
curves along with the displayed longitude of disc center and assumed axial inclination is
provided in the form of six numbers which appear below each of the reconstructed 
surface images we present. From left to right, 
these are as follows, using the values for
the image at upper left in Figure \ref{fig:Group2} as examples:
\begin{itemize}
\item ``190'' indicates that the longitude of disc center for the surface as
displayed is $\phi = 190^\circ$. This is not to be confused with the
actual sub-Earth longitude on the stellar surface, which varies from observation 
to observation. Unless otherwise indicated, the latitude of disc center for
all surface images is $\theta = +30^\circ$, as opposed to the actual 
sub-Earth latitude of $90^\circ - \alpha$.

\item ``45'' indicates that $\alpha = 45^\circ$ for the reconstruction.

\item ``20'' and ``23'' are the number of data points in the $B$ and $V$ light
curves used for the reconstruction, respectively.

\item ``268'' is the ``effective rms noise'' expressed in magnitudes (see 
\S \ref{sec:DataAnalysis}) for the $B$-filter light curve, multiplied
by $10^4$, so that the actual effective rms noise is 0.0268~mag in this case.
Similarly, ``326'' indicates that the effective rms noise for the $V$-filter
light curve is 0.0326~mag. Since the $B$-filter data were of lower noise in this
case, $B$ was designated the ``primary filter''  $n = 1$ for this inversion in 
the sense of the discussion in \S \ref{sec:LI_Algorithm}. Note that the
effective rms noise used in the inversion differs from actual deviation $\sigma=0.0311$~mag
reported in Figure \ref{fig:Group2Plots} for the $B$ light curve of Set 3. The reason for this
is discussed at the end of \S \ref{sec:DataAnalysis}.

\item ``2.7'' indicates that light curve data used to create the images span
2.7 revolution periods of the binary, which we assume also to be the rotation
period of the star because of tidal locking. (Note that if there is differential rotation,
this rotation period can strictly speaking only apply to some particular latitude.)
\end{itemize}

Before discussing the images, it should be noted that simulations \citep{HarmonCrews2000} 
show that when 
two separate circular spots are close together, they may appear in the 
reconstructions as a single elongated spot because of limitations in the 
resolution of the method. Indeed, what appears to be a single large spot may in
fact be a magnetic active region having complex structure which cannot be
resolved. We will thus tend to use ``spot'' and ``active region'' somewhat
interchangeably, but tending to prefer the latter terminology for a more 
extensive dark region on an image with a decidedly non-circular appearance.
Furthermore, as mentioned above, ``bridges'' which appear to connect spots together
may be artifacts of the inversion procedure.

The Set 3 images for both $\alpha = 45^\circ$ and $\alpha = 60^\circ$ 
appear to show a lower-latitude spot at left and a higher-latitude spot at
right connected by a bridge. The persistence of spottedness at high latitudes 
near the upper center of each image suggests that there is indeed a high-latitude
magnetic active region in this location, and that the bridge is not simply an
inversion artifact in this case.

If II Peg exhibits differential rotation in the same sense as the Sun in that
lower latitudes have higher angular velocities, then we should see the longitude
difference between the lower-latitude spot and the other features diminish over
time. This does appear to happen as one peruses the images from left to right.
In particular, the lower-latitude spot appears to catch up to and move 
underneath the high-latitude active region. This could represent two separate
spots which appear merged in the images due to the limited resolution of the
reconstructions, or it could represent a single large active region extended in
latitude, which rotates (roughly speaking) about an axis through its center
due to shearing by differential rotation carrying its southern end around
the star faster than its northern end.

The interpretation of the spot appearing at far right in the 
Set 3 images is problematic because it clearly appears at 
a higher latitude than the spot on the left in this pair of images, but in the
Set 4 images the spot on the right appears to be at a 
comparable latitude to the one on the left. This may simply represent a
limitation in the latitude discrimination of the spots in the inversions,
which certainly isn't perfect, particularly given that only $B$ and $V$
light curves were inverted as opposed to having more filter data available.
The spot on the right 
appears in the $\alpha = 45^\circ$ image for Set 5, but
not in the $\alpha = 60^\circ$ image. In simulations it is seen that a small
spot close to a much larger spot may be ``masked'' by the larger spot, causing
the smaller spot not to appear in the reconstruction. That this spot appears
in the $\alpha = 60^\circ$ image suggests that such masking has occurred for the 
$\alpha = 45^\circ$ image. 

It is important to note that Sets 3 and 4 were 
acquired over a total span of 7.9 stellar rotations and the corresponding images differ 
noticeably, while Set 5 by itself spans 8.2 rotations. It is thus likely that there
was significant evolution of the spots during the time Set 5 was acquired. In general,
all the surface images
shown here should be regarded as representing a sort of ``average'' appearance
of the starspots during the interval over which the data were collected.

It should also be noted that only twelve $B$ and thirteen $V$ observations were
available for the Set 6 time span. Thus, some
additional caution should be used in interpreting the corresponding images.

\begin{figure}[!htbp]
\centering
\plottwo{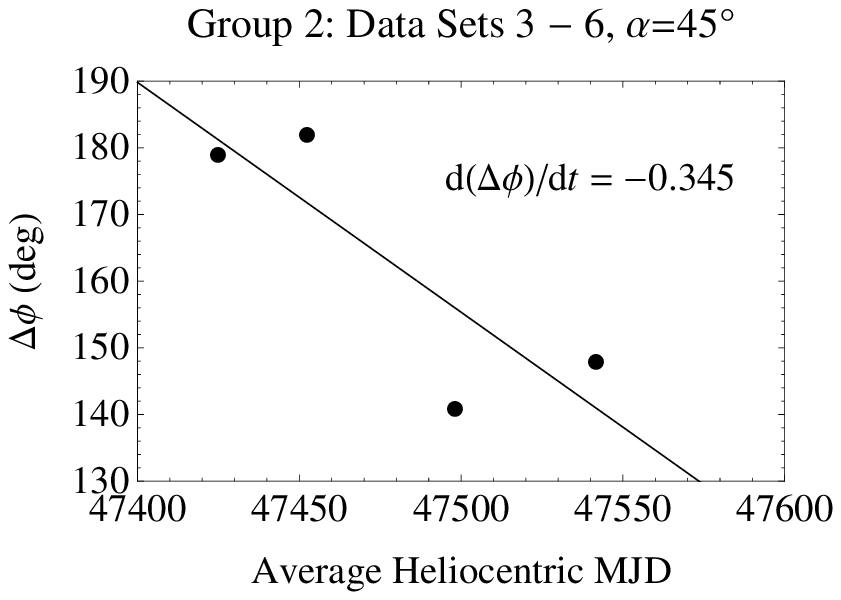}{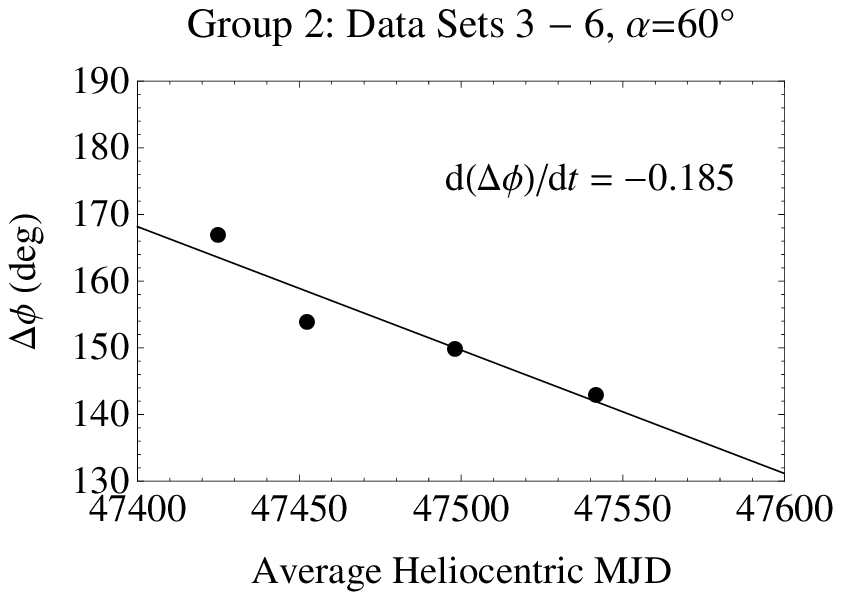}
\caption{\label{fig:Group2DiffRot} Plots of the span in longitude $\Delta
\phi$ of the active regions
shown in the inversions in Figure \ref{fig:Group2} versus the heliocentric
MJD of the midpoint of the time spanned by each data set, along
with the least-squares best-fit line. The method used to determine $\Delta \phi$
is detailed in the text. In this figure and in similar figures which follow it,
the slopes of the best fit lines expressed in units of degrees per day are indicated
on the plots.}
\end{figure}

Figure \ref{fig:Group2DiffRot} shows plots of the total span in longitude
$\Delta \phi$ of the spotted region versus the heliocentric Modified Julian Date of the
midpoint of the time interval over which the data were collected corresponding
to each image shown in Figure \ref{fig:Group2}. The plot on the left
is for $\alpha = 45^\circ$, while the one on the right is for 
$\alpha = 60^\circ$. To obtain $\Delta \phi$, an \emph{IDL}\footnote{\emph{IDL} is a
registered trademark of ITT Visual Information Solutions, Inc.} 
widget which displays the reconstructed image for user-adjusted values of the
latitude and longitude of disc center was employed. The stellar equator
was placed at disc center, and the longitude of disc center varied so as to
determine the longitudes of the easternmost and westernmost dark patches within
the spotted region. This is admittedly an imprecise determination, since there
is no sharp delineation between what constitutes a ``spot patch'' versus a 
``photosphere patch.'' Nevertheless, in practice the edges of the spotted
regions in the images are reasonably well-defined, as can be seen in Figure
\ref{fig:Group2} and similar figures below. A line was then fitted to
the data using a simple unweighted least-squares procedure. The slope of the
line represents a measure of the rate at which the longitude span of the 
spotted region is changing, expressed in degrees per day. There is considerable
variation
between the results for $\alpha = 45^\circ$ and $\alpha = 60^\circ$, but both
plots behave as would be expected if II Peg exhibits differential rotation
in the same sense as the Sun: $\Delta \phi$ decreases with time, because the
trailing region of the active region is at lower latitudes than is the leading 
region.

In \S \ref{sec:Discussion}, we show that the fits shown in 
Figure \ref{fig:Group2DiffRot} and similar figures for the other data
sets
showing evidence of differential rotation are consistent with the conclusions
of \citet{Henryetal1995} regarding the value of the differential rotation
coefficient $k$
defined in equation (\ref{eq:DiffRotP}).

\subsection{\label{sec:Group3} Group 3: MJD 47779.7480 -- 47926.5752}

\begin{figure}[!htbp]
\centering
\epsscale{0.75}
\plotone{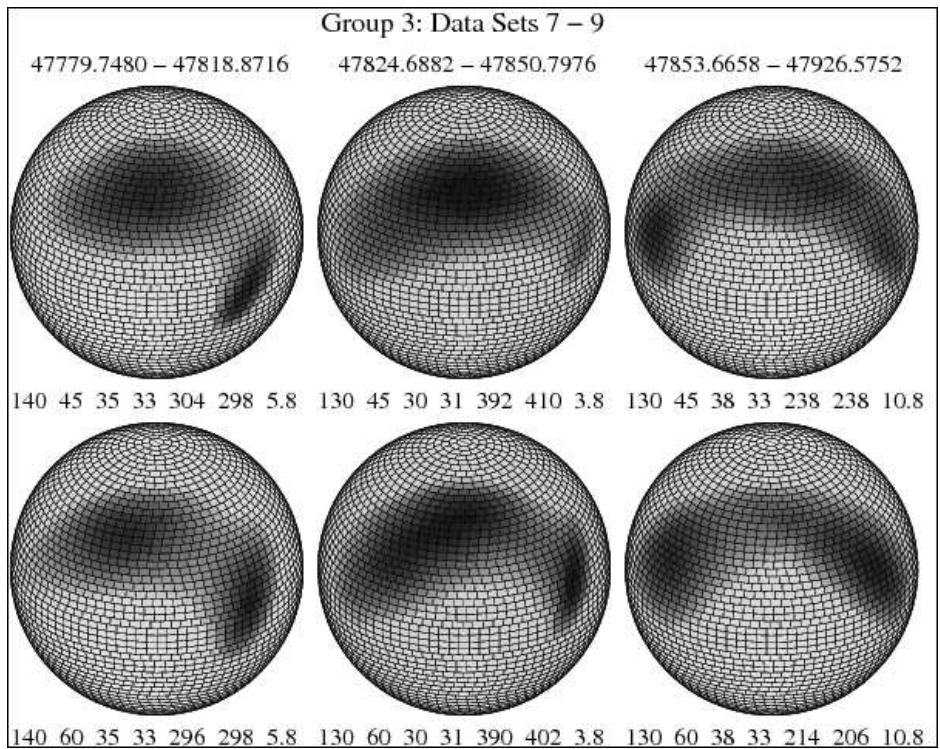}
\caption{\label{fig:Group3} Inversions of Group 3, consisting of three data sets spanning 
MJD 47779.7480 -- 47926.5752 (1989 September 10 -- 1990 February 4). 
Top row: $\alpha = 45^\circ$. Bottom row: $\alpha = 60^\circ$. }
\epsscale{1}
\end{figure}

Figure \ref{fig:Group3} shows inversions of Group 3, consisting of three data sets
obtained from MJD 47779.7480 -- 47926.5752
(1989 September 10 -- 1990 February 4), with $\alpha = 45^\circ$ in the top row
and $\alpha = 60^\circ$ in the bottom row. There are no light curve data for the
interval 1989 January 31 -- September 9. Thus, no information is available
regarding the evolution of the surface between the end of the time span covered
in
\S \ref{sec:Group2} and that covered here.

The images in Figure \ref{fig:Group3} give the distinct impression 
that the low-latitude spot on the right is moving away from higher-latitude
active regions to its right due to differential rotation, though it should be said that
the spot at far left in the third column appears to be a new feature which has
emerged within the interval covered by these light curves.

An interesting question is whether or not the low-latitude spot at right in the
images in Figure \ref{fig:Group3} is the same as the low-latitude spot
at left in the images in Figure \ref{fig:Group2}. If so, then over the 
combined span MJD 47417.7681 -- 47926.5752 (1988 September 11 -- 1990 February 4) 
we see the low-latitude spot
go from being well behind the high-latitude spot/active region in longitude to
being well ahead of it, and thus a distinctive manifestation of differential rotation.

The images are consistent with this hypothesis.
The midpoint of Data Set 7 corresponding
to the leftmost images in Figure \ref{fig:Group3} was at MJD 47798.8,
while the midpoint of Data Set 6 corresponding
to the rightmost images in Figure \ref{fig:Group2} was at MJD 47451.1, so
that 257.7 days elapsed between the two midpoints. Taking 
$|d(\Delta\phi)/dt| = 0.345\mbox{ deg d$^{-1}$}$ for $\alpha = 45^\circ$ from Figure
\ref{fig:Group2DiffRot} as a rough estimate of rate at which the
low-latitude spot should separate in longitude from the high-latitude spot/active region near
the central meridian of each figure, we would expect the longitude of the
low-latitude spot to have increased relative to the high-latitude spot by approximately 
$90^\circ$ in this time span. Using $|d(\Delta\phi)/dt| = 0.185\mbox{ deg
d$^{-1}$}$ for $\alpha = 60^\circ$ yields a relative increase in longitude of $48^\circ$ by
the low-longitude spot. The actual amount by which the low-latitude spot is ahead of the high-latitude spot in the images for Set 7 is about $60^\circ$. If the
images for Set 6 in Figure \ref{fig:Group2} are
taken to represent the two spots having approximately equal longitudes, then the
low-latitude spot being $60^\circ$ ahead in Set 7 is within the range of 
$48^\circ$ -- $90^\circ$ suggested by the plots in Figure \ref{fig:Group2DiffRot}.

\begin{figure}[!htbp]
\centering
\plottwo{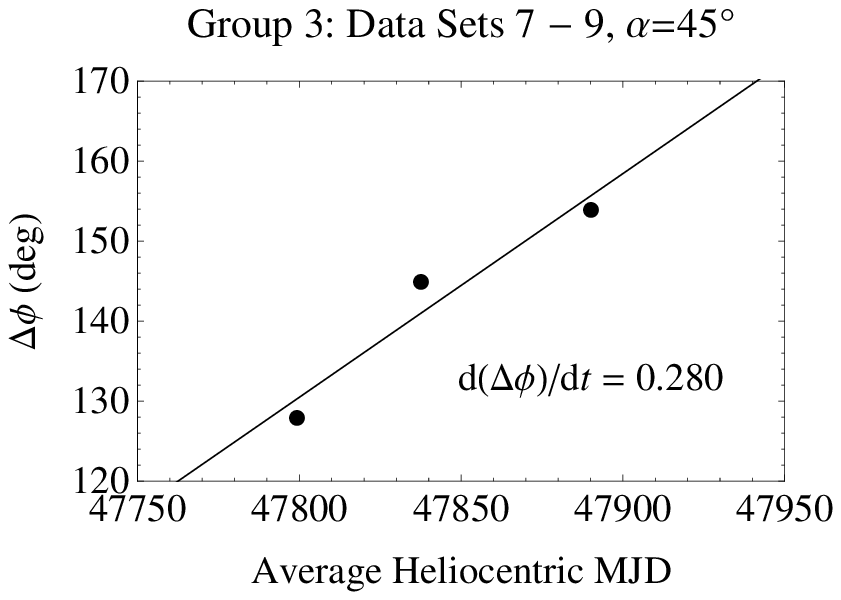}{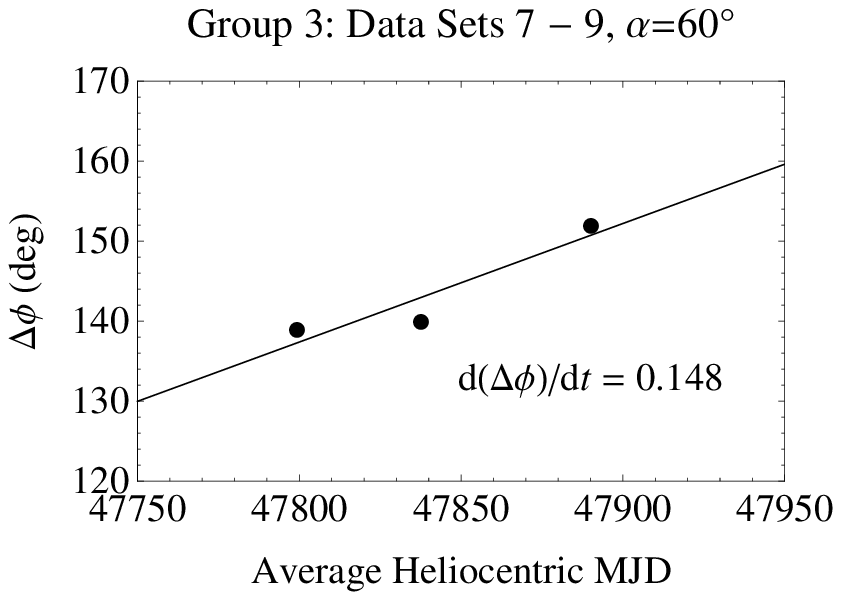}
\caption{\label{fig:Group3DiffRot} Span in longitude $\Delta \phi$ of the active regions
shown in the inversions in Figure \ref{fig:Group3} versus the MJD of the midpoint
of the time spanned by each data set.}
\end{figure}

Figure \ref{fig:Group3DiffRot} shows how $\Delta \phi$ for the
Group 3 data sets varies in time, obtained
via the same procedure as the plots in Figure \ref{fig:Group2DiffRot}. In
this case we expect $d(\Delta\phi)/dt > 0$ for solar-sense differential
rotation, since the low-latitude spot now starts out ahead in longitude. The magnitudes of
the slopes are comparable to those in the plots in Figure \ref{fig:Group2DiffRot},
strengthening the interpretation that we are seeing the effects of differential
rotation throughout this data set and that discussed in the preceding section.

\subsection{\label{sec:Group11} Group 11: MJD 50391.8348 -- 50856.5946}

\begin{figure}[!htbp]
\centering
\plotone{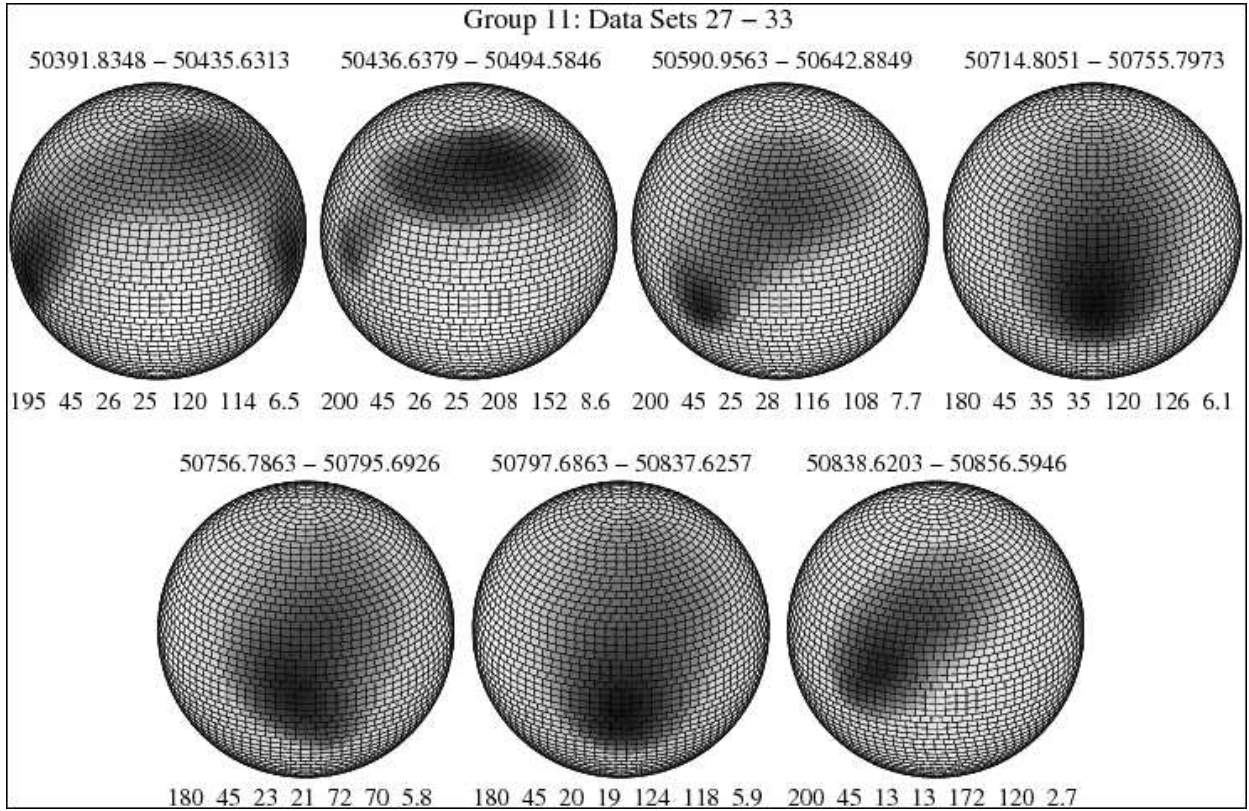}
\caption{\label{fig:Group11_45} Inversions of Group 11, consisting of seven data sets 
spanning MJD 50391.8348 -- 50856.5946
(1996 November 4 -- 1998 February 12). All images are for $\alpha = 45^\circ$. 
Note that the last image is from a perspective advanced $20^\circ$ in longitude relative 
to the preceding one. In addition, it was obtained from a sparse data set, as 
described in the text.}
\end{figure}

\begin{figure}[!htbp]
\centering
\plotone{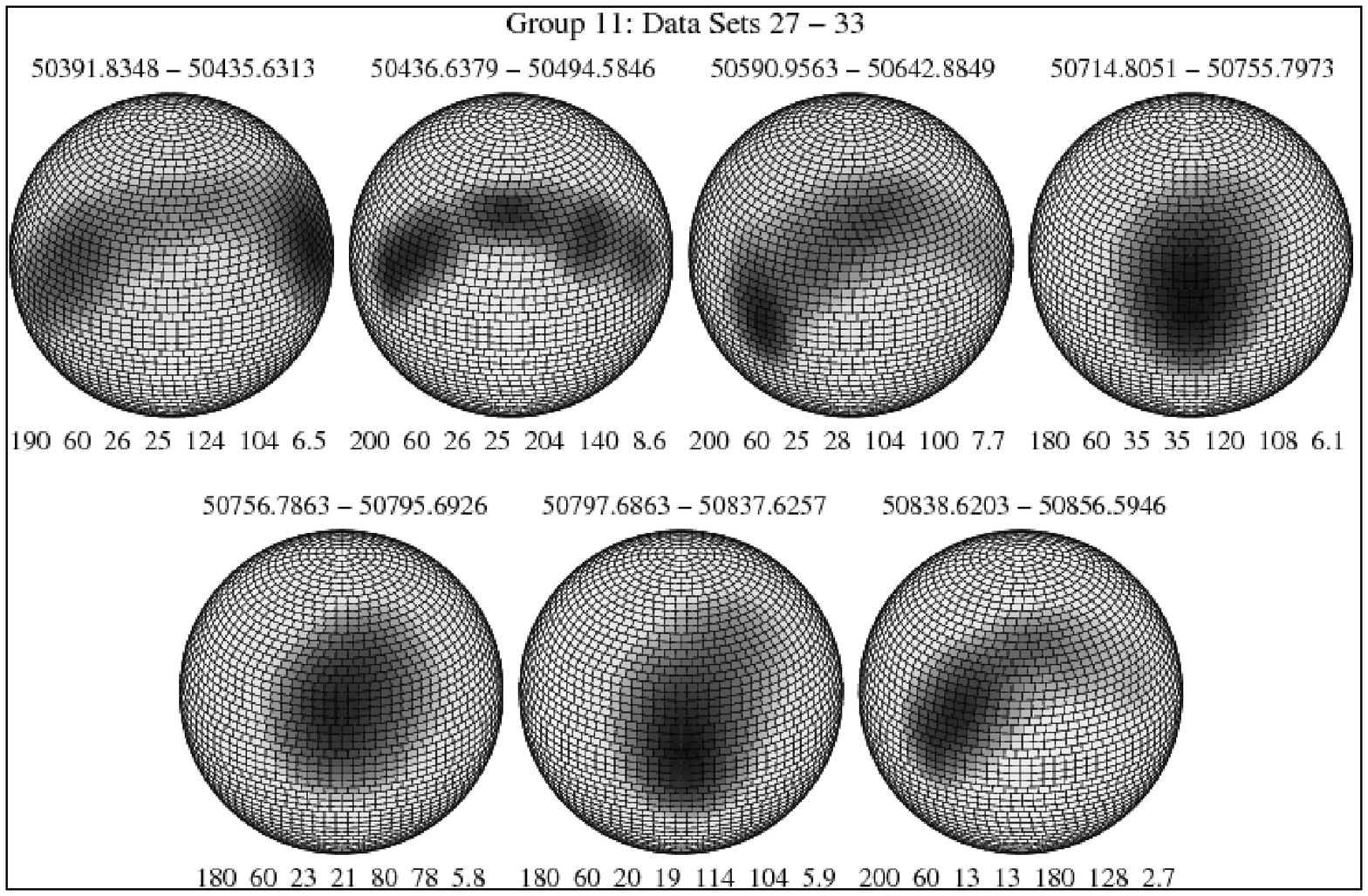}
\caption{\label{fig:Group11_60} Inversions of the same data sets as in
Figure \ref{fig:Group11_45}, for $\alpha = 60^\circ$. 
Note that the last image is from a perspective advanced $20^\circ$ in longitude relative 
to the preceding one. In addition, it was obtained from a sparse data set, as 
described in the text.}
\end{figure}

Figure \ref{fig:Group11_45} shows the inversions of Group 11, consisting of 
seven data sets obtained from MJD 50391.8348 -- 50856.5946 
(1996 November 4 -- 1998 February 12), all for an assumed
inclination 
of $\alpha = 45^\circ$. Figure \ref{fig:Group11_60} shows inversions of
the same light curves for $\alpha = 60^\circ$. The first 4 images (Sets 27 -- 30) in both figures
appear to show a trailing low-latitude spot at left catching up with and passing
under a high-latitude feature near the central meridian, highly suggestive of
differential rotation in the same sense as that of the Sun. 

From this we would expect to see the low-latitude spot begin to lead the
high-latitude feature and thus appear to the right of it in the next three images 
(Sets 31 -- 33). In the $\alpha = 
45^\circ$ image for Set 31, some hint that this is
occurring is seen in the way that the dark southern part of the spotted region
is tilted along a diagonal such that its southernmost extension is ahead in
longitude. This is not seen in the corresponding $\alpha = 60^\circ$ image. 

The Set 32 images for both inclinations are similar to the
Set 30 images. One may speculate that when previously 
separate large active regions are carried into proximity by differential rotation, 
they may merge or otherwise interact such
that magnetic tension resists deformation or separation by the differential rotation
of the surrounding plasma. If so, this would argue against our interpretation
that the southern spot in the data set described in \S \ref{sec:Group3}
is the same as that seen in the \ref{sec:Group2} data. However, it may be that 
whether or not this proposed suppression of differential rotation occurs depends on 
the details of the magnetic field configuration and the minimum separation
achieved.

The Set 33 images show high-latitude activity leading the southern portion of the 
spotted region in longitude. This gives the appearance of the sense of the
differential rotation having reversed, but it should be noted that both the $B$ and the $V$
light curves inverted to create these images had only 13 data points, which calls the 
reliability of these images into question. It is also possible that new high-latitude spot activity
had simply emerged. This interpretation
is supported by the appearance of the images for Set 32,
particularly for the $\alpha = 60^\circ$ inversions. High-latitude activity does seem
to be protruding at the northern end of the spotted region, which may represent
the beginning of the emergence of new magnetic flux.

\citet{Berdyugina1998b} present Doppler images for 1996 October and 1997 June,
August, and December, the first of which shortly precedes and the latter three of which
overlap in time with the photometric images discussed in this section.
We have no data for 1996 October, but their image gives some indication of the
high-latitude region seen in between the two initial spots in our later images. As is
typically the case, the spots seen in their Doppler images are at higher latitudes than in our
photometric images.

Their 1997 June image shows a large high-latitude active region with a large
$(\sim 150^\circ)$ extent in longitude, along with a pair of lower-latitude spots. 
In their 1997 August image, the extent in longitude of the large active region has diminished
considerably, and it appears that this is largely due to the disappearance of activity 
at the leading edge of the region. A smaller spot persists on the opposite side of the star. 
In their 1997 December image, the width in longitude of the large active region has further
diminished, but this time it appears that either the activity along the trailing edge has
diminished, or that the trailing 
edge has caught up to other parts of the region. There is a southward projection
at the trailing edge which is not present in the 1997 August image.

Our Set 29 (1997 May 22 -- July 13) images differ considerably from their 1997 June
image. Ours show a spot close to the equator trailing a higher-latitude active region. 
Our Set 27 (1996 November 4 -- December 18) images show the larger low-latitude spot present in
their 1997 June image, which appears to fade in our Set 28 (1996 December 19 -- 1997 February 15)
images. Our Set 30 (1997 September 23 -- November 3), Set 31 (1997 November 4 -- December 13),
and Set 32 (1997 December 15 -- 1998 January 24) images  are comparable to their 
1997 December image in showing a single large active 
region on one side of the star. The southern protrusion in their image could
explain the fact that our images show activity down to near the equator, though in their
image the protrusion only extends down to $\sim 35^\circ$ latitude. 

\begin{figure}[!htbp]
\centering
\epsscale{1}
\plottwo{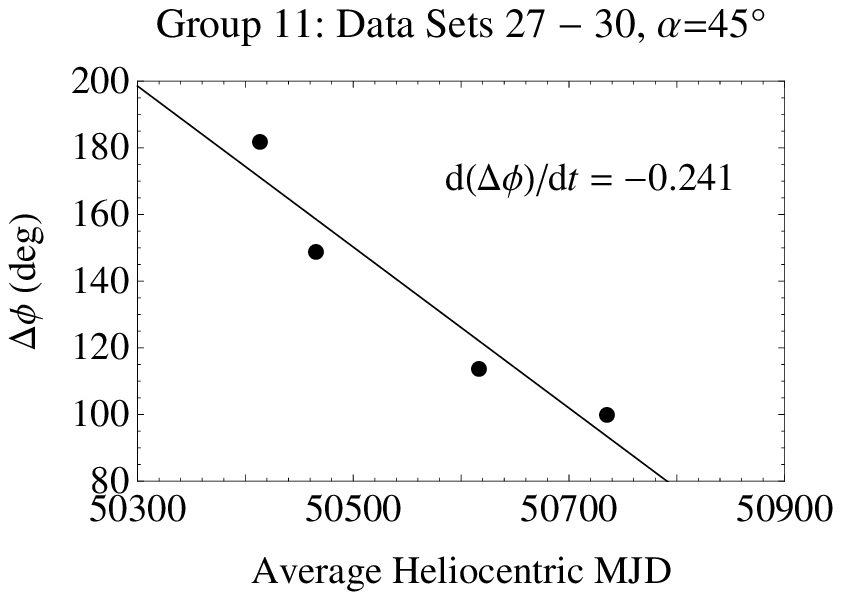}{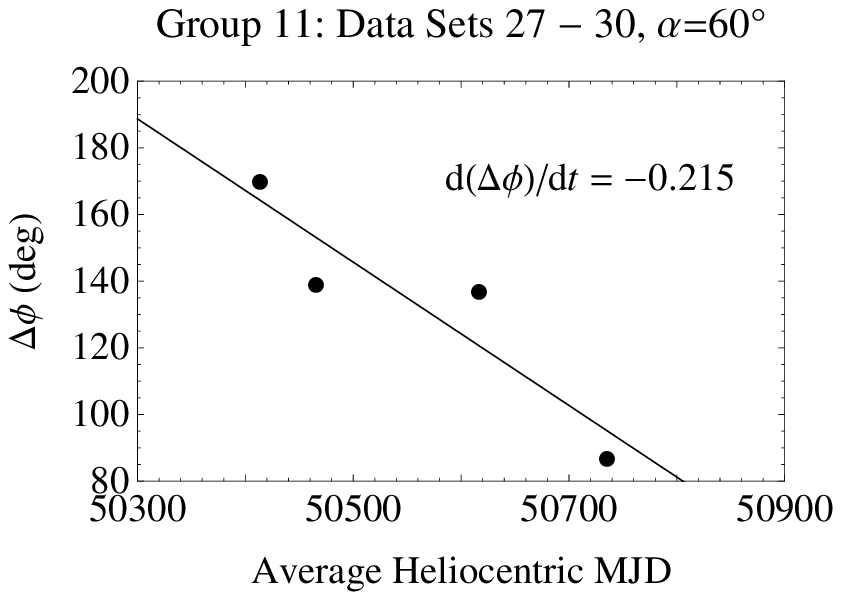}
\vspace{\baselineskip}
\plottwo{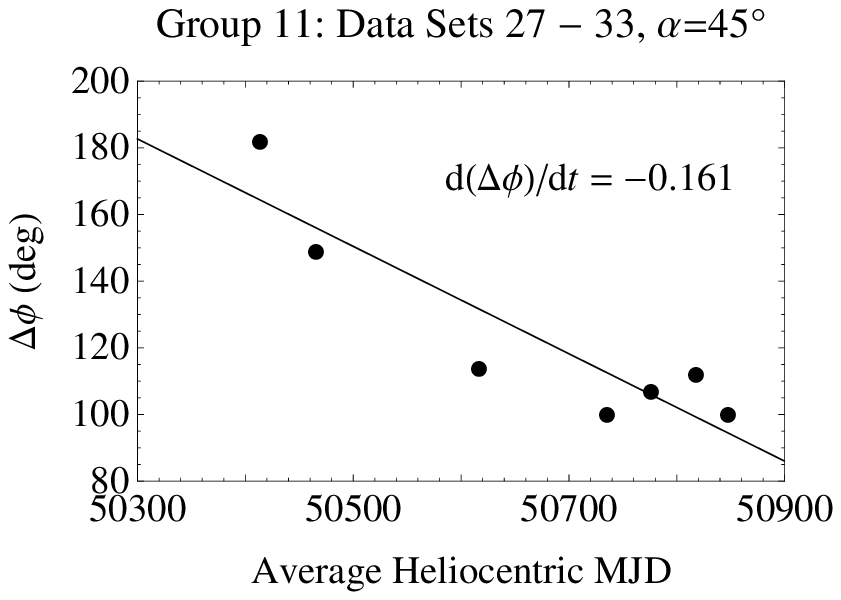}{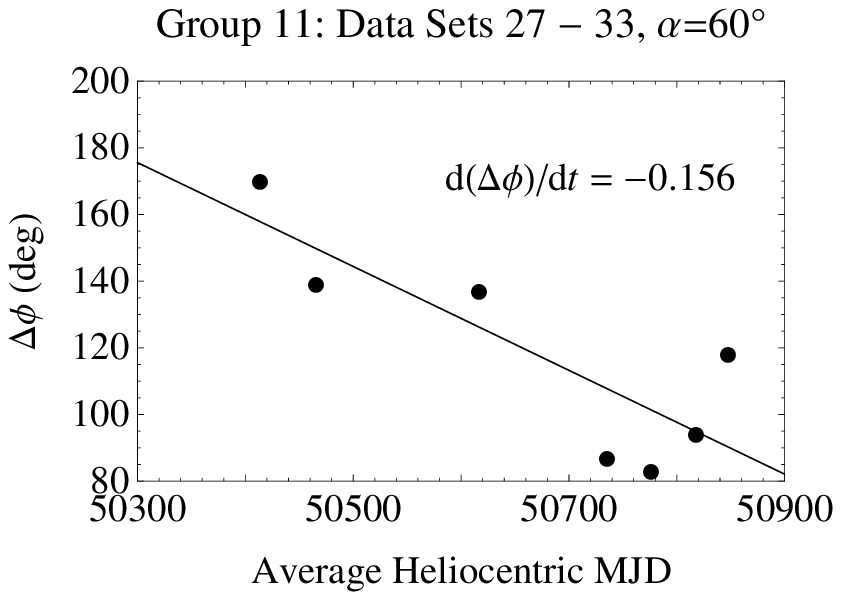}
\caption{\label{fig:Group11DiffRot} Span in longitude $\Delta \phi$ of the
active regions
shown in the inversions in Figures \ref{fig:Group11_45} and
\ref{fig:Group11_60}
versus the MJD of the midpoint of the time spanned by
each data set. For the plots at top, only the results for Sets 27 -- 30 are included, 
while for the plots at bottom, the results for all seven data sets in Group 11 are shown.}
\end{figure}

Figure \ref{fig:Group11DiffRot} shows plots of the variation in
$\Delta\phi$ with time for the seven data sets in Group 11.
The two plots at top include only the results for the first four data
sets for each inclination, while the two at bottom include all seven sets. 
The fits are better for the former two since as discussed above, the trend toward
narrowing of the spotted region in longitude ceases for the latter three data sets.

\subsection{\label{sec:Group12} Group 12: MJD 51085.9791 -- 51224.5942}

\begin{figure}[!htbp]
\centering
\plotone{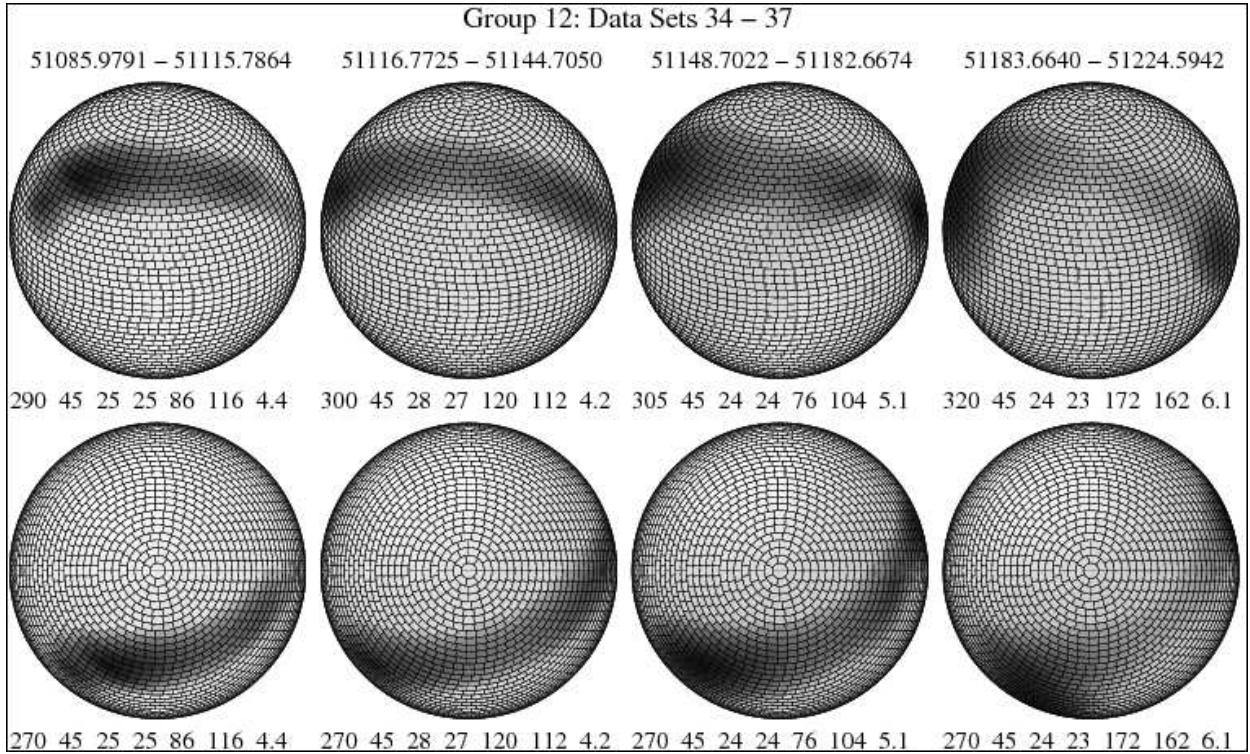}
\caption{\label{fig:Group12_45} Inversions of Group 12, composed of four data sets spanning 
MJD 51085.9791 -- 51224.5942 (1998 September 29 -- 1999 February 15). All images are for 
$\alpha = 45^\circ$. Top row: latitude of disc center is $\theta = 30^\circ$. 
Bottom row: looking down on north pole.}
\end{figure}

\begin{figure}[!htbp]
\centering
\plotone{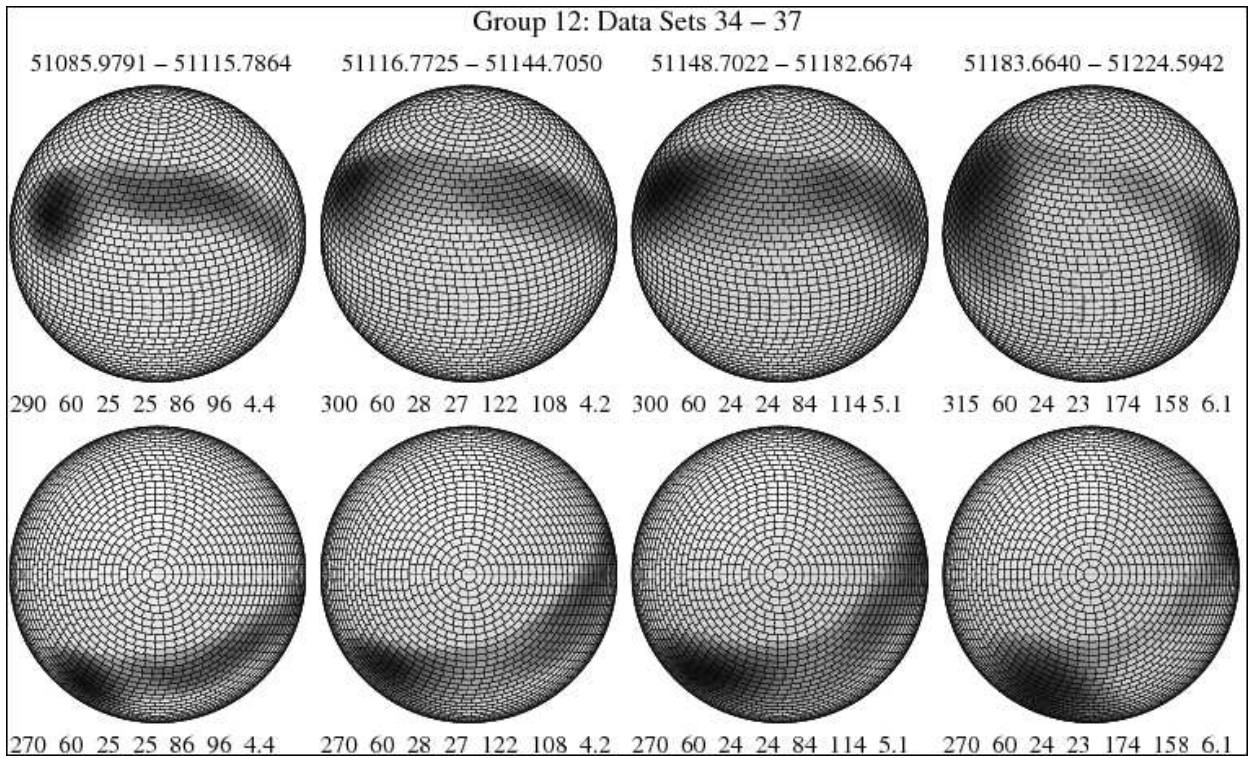}
\caption{\label{fig:Group12_60} Inversions of the same data sets as in
Figure \ref{fig:Group12_45}, for $\alpha = 60^\circ$.}
\end{figure}

Figure \ref{fig:Group12_45} shows inversions for $\alpha = 45^\circ$ of Group 12, 
consisting of four data sets obtained from MJD 51085.9791 -- 51224.5942 
(1998 September 29 -- 1999 February 15), while
Figure \ref{fig:Group12_60} shows the same for $\alpha = 60^\circ$. In both
figures, the images in the top row have latitude $\theta = 30^\circ$ at disc
center, while for the images in the bottom row the north (visible) rotation pole is at
disc center. 

For both assumed inclinations, the images show a leading low-latitude spot or
active region at right and a trailing high-latitude spot/active region at left.
As time progresses, the separation in longitude between these features
increases, just as would be expected if solar-sense differential rotation is carrying the 
low-latitude feature around the star faster than the high-latitude region. This 
is most easily seen in the images looking down on the pole. 

Note that the degree of magnetic activity appears to increase on both ends of
the spotted region during this time span.

\citet{Berdyugina1999} present Doppler images for 1998 October and November
which overlap the interval under consideration here. Their 1998 October image shows an active
region spread extensively in longitude that is not resolved into separate spots; in the
November image, this region appears split into three separate spots. The difference is
conceivably due to the variation in the noise artifacts between the two images. 
Our Data Set 35 (1998 October 30 -- November 27) 
image shows a similar elongated feature, but lacks the southward protrusion seen
in the middle region of the active region in the Doppler images.

\begin{figure}[!htbp]
\centering
\plottwo{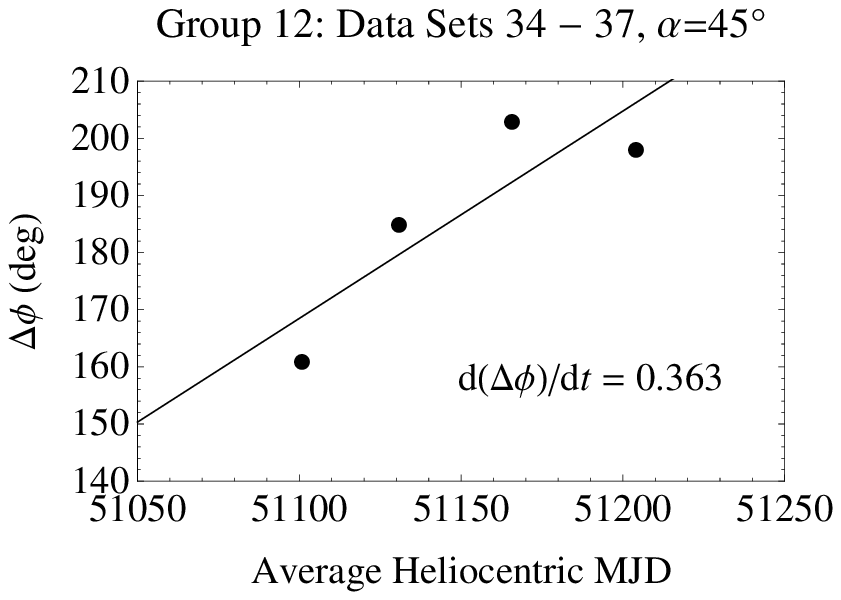}{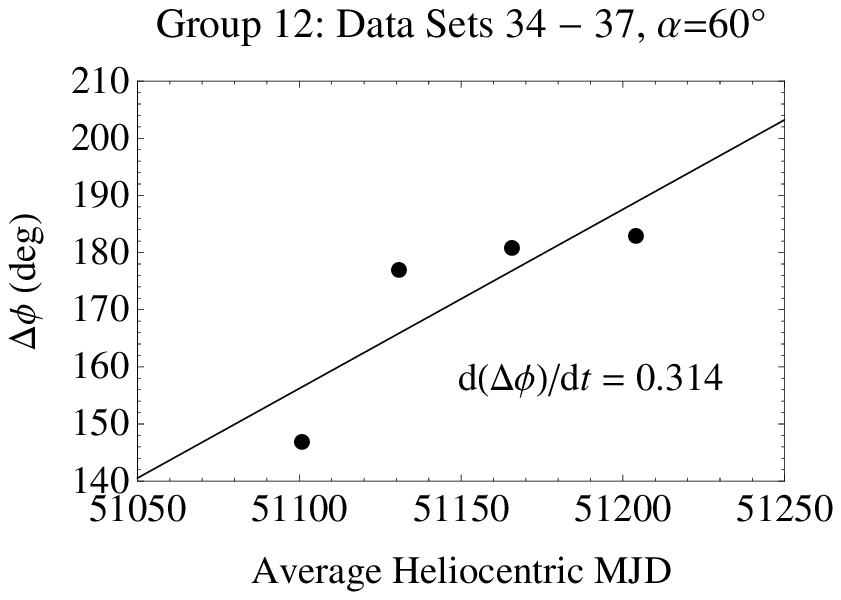}
\caption{\label{fig:Group12DiffRot} Span in longitude $\Delta \phi$ of the
active regions shown in the inversions in Figures \ref{fig:Group12_45} and
\ref{fig:Group12_60} versus the MJD of the midpoint of the time spanned by each data set.}
\end{figure}

Figure \ref{fig:Group12DiffRot} shows the variation of $\Delta\phi$ with
time for this time interval. As expected, $\Delta\phi$ increases with time, since the
low-latitude, faster-moving activity is leading in longitude.

\subsection{\label{sec:Group13} Group 13: MJD 51429.8915 -- 51586.5994}

\begin{figure}[!htbp]
\centering
\plotone{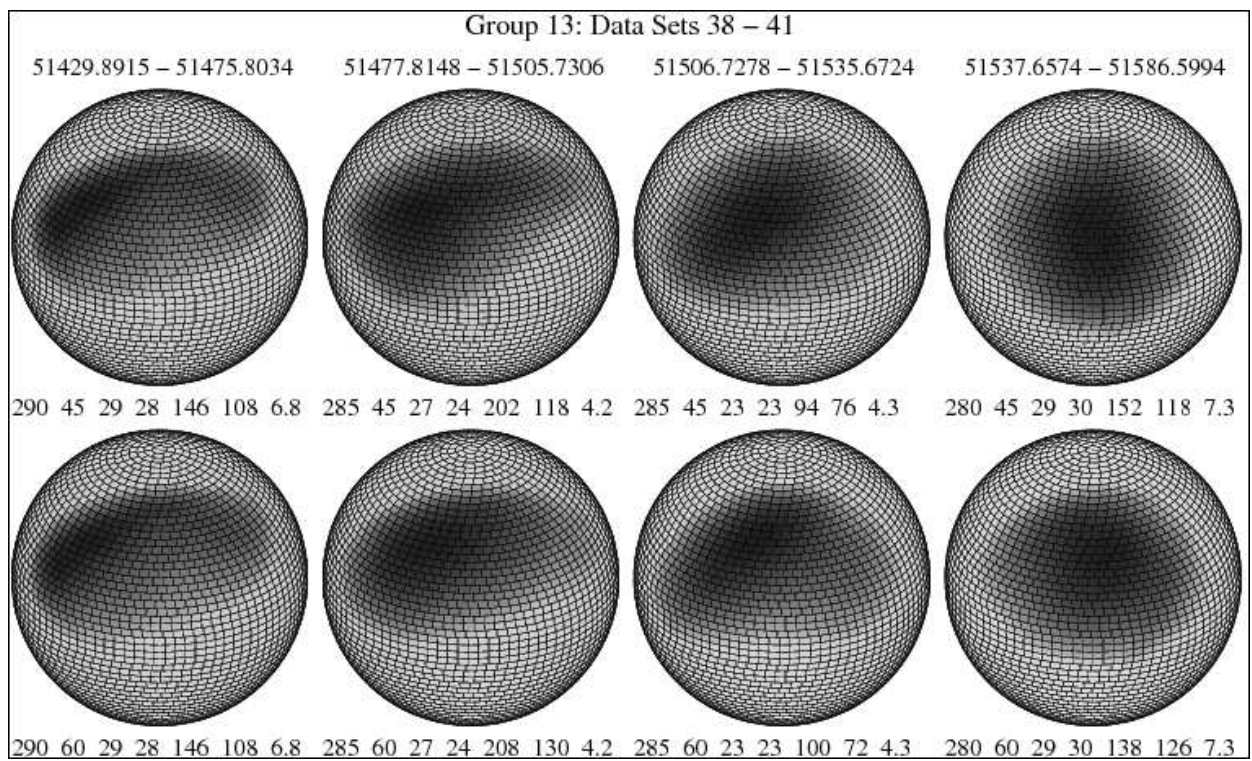}
\caption{\label{fig:Group13} Inversions of Group 13, consisting of four data sets 
spanning MJD 51429.8915 -- 51586.5994 (1999 September 8 -- 2000 February 12). 
Top row: $\alpha = 45^\circ$. Bottom row: $\alpha = 60^\circ$. }
\end{figure}

Figure \ref{fig:Group13} shows inversions of Group 13, composed of 
four light curves obtained between MJD 51429.8915 -- 51586.5994
(1999 September 8 -- 2000 February 12), with $\alpha = 45^\circ$ for the top row
and $\alpha = 60^\circ$ for the bottom row. For both assumed inclinations, the
trailing edge of the spotted region appears to be at lower latitudes than the
leading edge. This may represent two separate spots which are too close to be
resolved, or two spots embedded within an active region.

For differential rotation in the same sense as the Sun, we expect
that over time the separation in longitude between the leading and trailing
edges (or the two spots) should diminish. This is what is seen. The last pair of
images appear to show that the low-latitude trailing spot (or extension of the spotted
region) has caught and passed underneath higher-latitude activity.  

\citet{Gu2003} produced surface maps of II Peg via Doppler imaging for 1999 July
-- August, 2000 February and 2001 November -- December, the first two of which are relevant
to the present discussion. Their 1999 July -- August image is qualitatively similar to
our Set 38 (1999 September 8 -- October 24) images, showing a broad region of high-latitude
activity with southward projections at its leading and trailing ends, with the trailing end
projecting further south. Their 2000 February image primarily gives the appearance 
that the active region may have simply shrunk in longitude. However, it could
also be argued that the trailing edge has caught up to the leading edge.

\begin{figure}[!htbp]
\centering
\plottwo{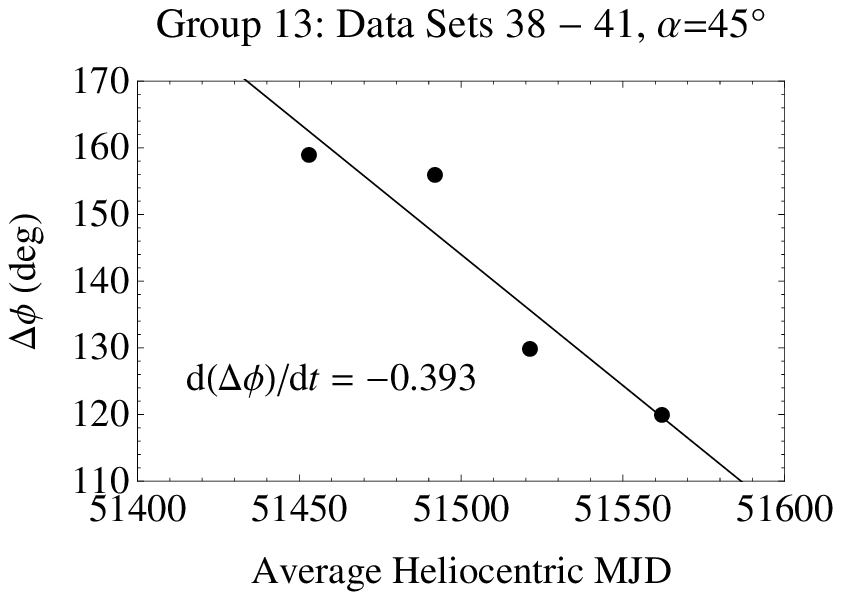}{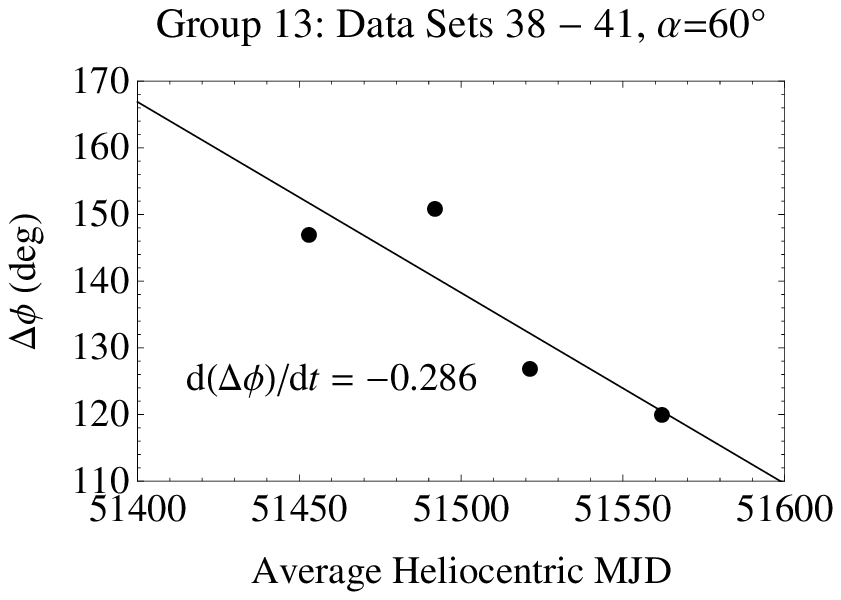}
\caption{\label{fig:Group13DiffRot} Span in longitude $\Delta \phi$ of the
active regions shown in the inversions in Figure \ref{fig:Group13} versus the
MJD of the midpoint of the time spanned by each data set.}
\end{figure}

Figure \ref{fig:Group13DiffRot} shows that the
the extent in longitude $\Delta \phi$ of the spotted region described here
diminishes with time, as expected for solar-sense differential rotation.

\subsection{\label{sec:Group14} Group 14: MJD 51805.8758 -- 51946.6051}

\begin{figure}[!htbp]
\centering
\epsscale{0.75}
\plotone{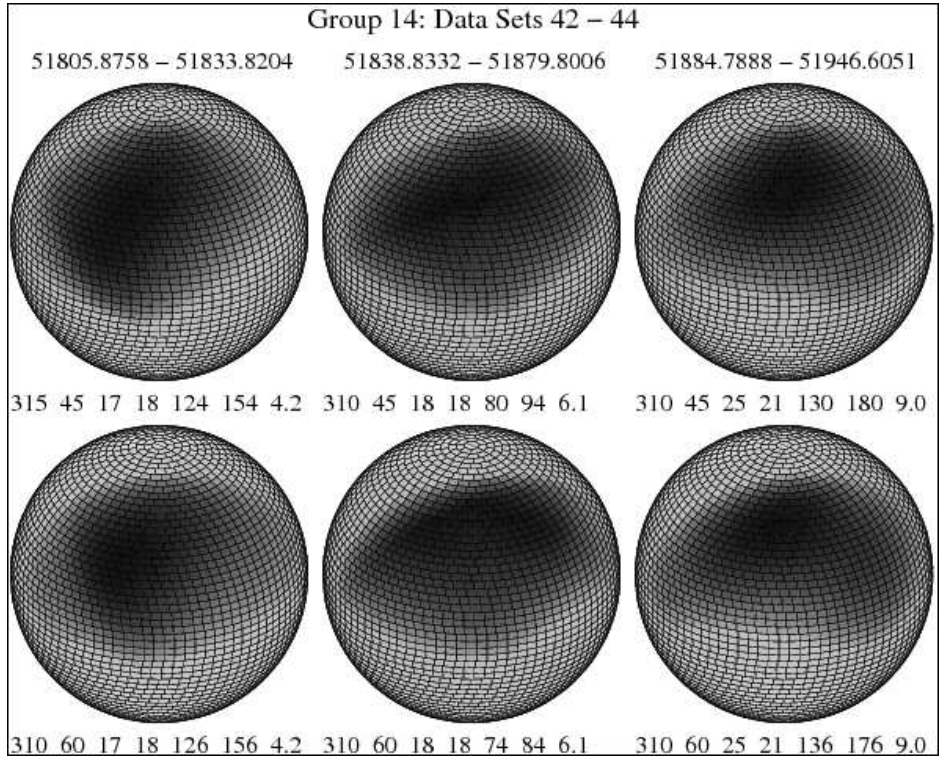}
\caption{\label{fig:Group14} Inversions of Group 14, composed of three data sets spanning 
MJD 51805.8758 -- 51946.6051 (2000 September 18 -- 2001 February 6). 
Top row: $\alpha = 45^\circ$. Bottom row: $\alpha = 60^\circ$.}
\end{figure}

The final time interval over which we saw good evidence for differential
rotation was from MJD 51805.8758 -- 51946.6051 (2000 September 18 -- 2001 February 6),
which we partitioned into three data sets which were assigned to Group 14.
Figure \ref{fig:Group14} shows the corresponding inversions, with assumed inclination 
$\alpha = 45^\circ$ for top row and $\alpha = 60^\circ$ for the bottom row.

For both values of $\alpha$, a protrusion toward lower latitudes can be seen at the
bottom of the spotted region. This protrusion appears to drift toward greater
longitudes over the time interval covered by these observations, suggesting that it is a
spot being carried around the equator faster than the rest of the spotted region
because of differential rotation. 

However, we cannot quantify this differential rotation by plotting the variation of
$\Delta \phi$ with time as for the data groups discussed above, because in this case
the total extent in longitude of the active region did not change significantly. The
easternmost and westernmost fringes of the active region appear to lie at similar middle
latitudes, so that we would not expect $\Delta \phi$ to change due to differential rotation
here. The longitude of the southward protrusion is within the range of longitudes
spanned by the activity to its north, so that its drift in longitude relative to the
rest of the spotted region is not manifested in the variation of $\Delta\phi$.

\begin{figure}[!htbp]
\centering
\epsscale{0.5}
\plotone{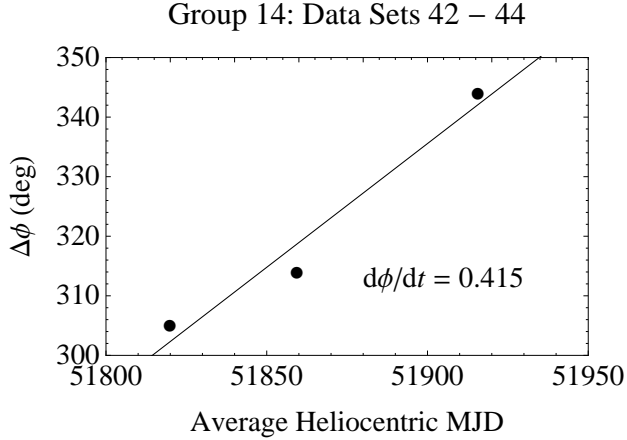}
\caption{\label{fig:Group14DiffRot} Longitude of the middle of the
southernmost protrusion of the spotted region for the Group 14 images in Figure
\ref{fig:Group14}.}
\end{figure}

Instead, in Figure \ref{fig:Group14DiffRot} we show the variation in
longitude of the middle of the southern protrusion with time. This is admittedly a vague 
conception due to the amorphous nature of the protrusion in these images, which
is likely just an artifact of the limitations in the resolution of the inversions. The 
procedure was to use the same \emph{IDL} widget used to obtain $\Delta\phi$
for other data sets to estimate the longitude of the midpoint of the southern
boundary of the protrusion. The same results were obtained for both $\alpha = 45^\circ$ and
$\alpha = 60^\circ$. As can be seen in Figure \ref{fig:Group14DiffRot}, the
longitude of the protrusion (or spot) increases approximately linearly with time, as would
be expected from differential rotation if the protrusion or spot maintains an
essentially fixed latitude.

\section{\label{sec:Discussion} Discussion}

In this study, we obtained images of the surface of II Pegasi by inverting data sets 
consisting of light curves collected over intervals of several months to look 
for evidence of differential rotation in the changes in the spot configurations. 
By contrast, \citet{Henryetal1995} inferred the presence of differential rotation using 
the very different approach of determining the rotation periods
of individual spots observed over spans of several years. It is of interest to
ascertain whether or not the results presented here are consistent with the earlier study.

In contrast with the relation used by Henry et al.\ given by equation
(\ref{eq:kf}), in the present study it is more natural to characterize the differential
rotation in terms of the rate of change in the difference in longitude between spots at different
latitudes in our images. 

If for simplicity we express longitudes in a non-rotating frame of reference (rather than in the
co-rotating frame used in the rest of this work), the longitude as a function of
time $t$ for a spot having latitude $\theta$ is just
\begin{equation}
\phi(t) = \phi_0 + t\Omega(\theta),
\label{eq:SpotPhi}
\end{equation}
where $\phi_0$ is the longitude of the spot at time $t = 0$, and  
$\Omega(\theta) = 2\pi/P(\theta)$ is the angular rotation frequency for latitude
$\theta$, with $P(\theta)$ being the corresponding rotational period.
From this the difference in longitude $\Delta\phi = \phi_2(t) -
\phi_1(t)$ of 
two spots at latitudes $\theta_1$ and $\theta_2$ is simply
\begin{equation}
\Delta\phi = \Delta\phi_0 + t\Delta\Omega,
\label{eq:SpotDeltaPhi}
\end{equation}
where $\Delta\phi_0$ is the separation at $t = 0$ and $\Delta\Omega =
\Omega(\theta_2) -
\Omega(\theta_1)$. Thus, a plot of $\Delta\phi$ versus $t$ will have slope
$\Delta\Omega$.

In terms of the rotational angular frequency, equation (\ref{eq:DiffRotP})
becomes
\begin{equation}
\Omega(\theta) = \Omega_\mathrm{eq}(1 - k\sin^2\theta),
\label{eq:DiffRotOmega}
\end{equation}
so that
\begin{equation}
k = \frac{\Delta\Omega}{\Omega_\mathrm{eq}(\sin^2\theta_1 - \sin^2\theta_2)}.
\label{eq:kFromDeltaOmega}
\end{equation}
In principle, then, if we observe two spots at known latitudes and plot
$\Delta\phi$
versus $t$ to find the slope $\Delta\Omega$, we can find $k$ from equation
(\ref{eq:kFromDeltaOmega}).

In practice, this is not possible because we cannot reliably determine precise spot latitudes
from our photometric inversions. However, we can use the plots in Figures 
\ref{fig:Group2DiffRot}, \ref{fig:Group3DiffRot}, \ref{fig:Group11DiffRot},
\ref{fig:Group12DiffRot} and \ref{fig:Group13DiffRot} to make reasonable estimates
of the value of $k$. In these figures, $\Delta\phi$
represents the extent in longitude of an active region rather than the longitude difference between
two spots, because the former could be more reliably estimated from our images. 

Using the revolution period $P = 6.724333\mbox{ d}$ of the binary as our estimate of 
$P_\mathrm{eq}$, we find $\Omega_\mathrm{eq} = 0.934395\mbox{ d}^{-1}$. Excluding the plots
having seven data points in Figure \ref{fig:Group11DiffRot}, the slopes 
$\Delta\Omega$ obtained from the plots for $\alpha = 45^\circ$ range from 
$\Delta\Omega = 0.24\mbox{ deg d$^{-1}$}$ to $0.39\mbox{ deg d$^{-1}$}$, with a mean of
$0.32\mbox{ deg d$^{-1}$}$. If we assume that $\theta_1 = 90^\circ$ and $\theta_2 = 0^\circ$
(which is almost certainly an overestimate of the spread in latitude responsible for the change 
in $\Delta\phi$, particularly given the results of Doppler imaging), we obtain a minimum
value for $k$ of $0.0045$, a maximum value of $0.0073$, and a mean value of $0.0059$.
For $\alpha = 60^\circ$, $\Delta\Omega$ from the slopes of the plots (again excluding the
seven-data-point plot in Figure \ref{fig:Group11DiffRot}) 
ranges from $0.15\mbox{ deg d$^{-1}$}$ to $0.31\mbox{ deg d$^{-1}$}$ with a mean
of $0.24\mbox{ deg d$^{-1}$}$. These yield a minimum value for $k$ of $0.0028$, a maximum value of $0.0059$, and a mean
of $0.0045$. These results accord well with the value $k = 0.005 \pm 0.001$ obtained 
for II Peg by Henry at al., which lends credence to our assertion that our images
represent genuine demonstrations of differential rotation, and to the assertion of
Henry et al.\ that stretching of active regions by differential rotation accounts for
features of the light curve of II Peg not explainable by their simple two-spot model.

Of course, the values obtained for $k$ are increased if the span of latitude is not
from pole to equator. The Doppler images of \citet{Berdyugina1998b,Berdyugina1999}
show activity extending from mid-to-high latitudes, but not including any polar spots,
so it seems reasonable to use $\theta_1 = 80^\circ$ and $\theta_2 = 45^\circ$ in
our estimates. For $\alpha = 45^\circ$, this results in minimum, maximum and mean values for
$k$ of $0.0096$, $0.0156$, and $0.0126$, respectively. For $\alpha = 60^\circ$ the corresponding
values are $0.0059$, $0.0125$, and $0.0095$. These results do not agree as well with the result
of Henry et al., but are certainly of the same order. It should also be noted that Henry et al.\
assumed that the spots they observed spanned a range in latitude from pole to equator, so
that they too would have obtained a larger value of $k$ had they assumed a smaller range.

\acknowledgements
N.V. acknowledges support from the Ohio Wesleyan University Summer
Science Research Program. G.W.H. acknowledges support from NASA, NSF, Tennessee State
University, and the State of Tennessee through its Centers of
Excellence program.


\begin{thebibliography}{}
\bibitem[Berdyugina \& Tuominen(1998)]{BerdyuginaTuominen1998} Berdyugina, S.\ V., 
\& Tuominen I.\ 1998, \aap, 336, L25

\bibitem[Berdyugina et al.(1998a)]{Berdyugina1998a} Berdyugina, S.\ V., Jankov, S., 
Ilyin, I., Tuominen, I., \& Fekel, F.\ C.\ 1998a, \aap, 334, 863

\bibitem[Berdyugina et al.(1998b)]{Berdyugina1998b} Berdyugina, S.\ V., Berdyugin, A.\ V.,
Ilyin, I., \& Tuominen, I.\ 1998b, \aap, 340, 437

\bibitem[Berdyugina et al.(1999)]{Berdyugina1999} Berdyugina, S.\ V., Berdyugin, A.\ V.,
Ilyin, I., \& Tuominen, I.\ 1999, \aap, 350, 626

\bibitem[Boyd et al.(1987)]{Boyd1987} Boyd, P.\ T., Garlow, K.\ R., Guinan, E.\ F.,
McCook, G.\ P., McMullin, J.\ P., \& Wacker, S.\ W.\ 1987, Inf.\ Bull.\ Var.\ Stars, No. 3089

\bibitem[Byrne \& Marang(1987)]{ByrneMarang1987} Byrne, P.\ B., \& Marang, F.\
1987, Irish Astron.\
J., 18, 84

\bibitem[Doyle et al.(1989)]{Doyle1989} Doyle, J.\ G., Butler, C.\ J., 
Byrne, P.\ B., Rodon\`{o}, M., Swank, J., \& Fowles, W.\ 1989, \aap, 223, 219

\bibitem[Chugainov(1976)]{Chugainov1976} Chugainov, P.\ F.\ 1976, Iz.\ Kry.\, 57, 31

\bibitem[Craig \& Brown(1986)]{CraigBrown1986}  Craig,
I.\ J.\ D., \& Brown, J.\ C.\ 1986, Inverse Problems in Astronomy (Boston:
A. Hilger)

\bibitem[Frasca et al.(2008)]{Frasca2008} Frasca, A., Biazzo, K., Ta\c{s},
G., Evren, S.\ \& Lanzafame, A.\ C.\ 2008, \aap, 479, 557

\bibitem[Gray(1992)]{Gray1992} Gray, D.\ F.\ 1992, The Observation and
Analysis of Stellar Photospheres (Cambridge: Cambridge University Press)

\bibitem[Gu et al.(2003)]{Gu2003} Gu, S.-H., Tan, H.-S., Wang, X.-B., \& Shan, H.-G.\ 2003,
\aap, 405, 763

\bibitem[Hall \& Henry(1994)]{HallHenry1994} Hall, D.\ S., \& Henry, G.\ W.\ 1994, 
IAPPP Comm., 55, 51

\bibitem[Harmon \& Crews(2000)]{HarmonCrews2000} Harmon, R.\ O., \& Crews, L.\ J.\
2000, \aj, 120, 3274

\bibitem[Henry(1995a)]{Henry1995a} Henry, G.\ W.\ 1995, in Robotic Telescopes: 
Current Capabilities, Present Developments, and Future Prospects for Automated Astronomy, 
ASP Conf.\ Ser.\ Vol.\ 79, eds. G. W. Henry and J. A. Eaton 
(Provo: Astronomical Society of the Pacific), p.\ 37

\bibitem[Henry(1995b)]{Henry1995b} Henry, G.\ W.\ 1995, in Robotic Telescopes: 
Current Capabilities, Present Developments, and Future Prospects for Automated Astronomy, 
ASP Conf.\ Ser.\ Vol.\ 79, eds. G. W. Henry and J. A. Eaton 
(Provo: Astronomical Society of the Pacific), p.\ 44

\bibitem[Henry et al.(1995)]{Henryetal1995} Henry, G.\ W., Eaton, J. A., Hamer, J., 
\& Hall, D.\ 1995, \apjs, 97, 513

\bibitem[Kurucz(1991)]{Kurucz1991} Kurucz, R.\ L.\ 1991, Harvard Preprint 3348

\bibitem[Messina(2008)]{Messina2008} Messina, S.\ 2008, 480, 495

\bibitem[Nations \& Ramsey(1981)]{NationsRamsey1981} Nations, H.\ L., \& Ramsey,
L.\ W.\ 1981, \aj, 86, 433

\bibitem[O'Neal, Saar, \& Neff(1998)]{ONeal1998} O'Neal, D., Saar, S.\ H., \& Neff, J.\ E.\
1998, \apj, 501, L73

\bibitem[Poe \& Eaton(1985)]{PoeEaton1985} Poe, C.\ H., \& Eaton, J.\ A.\ 1985, \apj, 289, 644

\bibitem[Rodon\`{o} et al.(1986)]{Rodono1986} Rodon\`{o}, M.\ et al. 1986, \aap, 165, 135

\bibitem[Rodon\`{o} et al.(2000)]{Rodono2000} Rodon\`{o}, M., Messina, S., Lanza, A.\ F.,
Cutispoto, G., and Teriaca, L.\ 2000, \aap, 358, 624

\bibitem[Rucinski(1977)]{Rucinski1977} Rucisnski, S.\ M.\ 1977, \pasp, 89, 280

\bibitem[Thompson \& Craig(1992)]{ThompsonCraig1992} Thompson, A. M., \& Craig,
I.\ J.\ D.\ 1992, \aap, 262, 359

\bibitem[Turchin(1967)]{Turchin1967} Turchin, V.\ F.\ 1967, USSR
Comput.\ Math.\ and Math.\ Phys., 7, 79

\bibitem[Twomey(1977)]{Twomey1977} Twomey, S.\ 1977, Introduction to the
Mathematics of Inversion in Remote Sensing and Indirect Measurements
(Amsterdam: Elsevier)  

\bibitem[Van Hamme(1993)]{VanHamme1993} Van Hamme, W.\ 1993,
\aj, 106, 2096 

\bibitem[Vogt(1981a)]{Vogt1981a} Vogt, S.\ 1981a, \apj, 247, 975

\bibitem[Vogt(1981b)]{Vogt1981b} Vogt, S.\ 1981b, \apj, 250, 327

\bibitem[Vogt \& Penrod(1983)]{VogtPenrod1983} Vogt, S.\ S., \& Penrod, G.\ D.\ 1983,
\pasp, 95, 565

\bibitem[Wild(1989)]{Wild1989} Wild, W.\ J.\ 1989, \pasp, 101, 844
\end{thebibliography}
\end{document}